\documentclass[useAMS,usenatbib]{mn2e}

\usepackage[pdftex]{graphicx}
\usepackage{amsmath}

\newcommand{\kms}{{km s$^{-1}$}}

\usepackage{relsize}

\newcommand{\hi}{%
  \relax
  \ifmmode
    \textrm{\textsc{HI}}
  \else
    \textsc{H{\smaller}I}
  \fi
}

\newcommand{\simgt}{\lower.5ex\hbox{$\; \buildrel > \over \sim \;$}}
\newcommand{\simlt}{\lower.5ex\hbox{$\; \buildrel < \over \sim \;$}}

\title[The Shape of Dark Matter Haloes V.]{The Shape of Dark Matter Haloes\\
V. Analysis of observations of edge-on galaxies}
\author[S. P. C. Peters et al.]{S. P. C. Peters$^{1}$,
P. C. van der Kruit$^{1}$\thanks{For more information, please contact P.C. van der Kruit by email at vdkruit@astro.rug.nl.}, 
R. J. Allen$^{2}$ and K. C. Freeman$^{3}$\\
$^{1}$Kapteyn Astronomical Institute, University of Groningen, P.O.Box 800, 9700AV Groningen, the Netherlands\\
$^{2}$Space Telescope Science Institute, 3700 San Martin Drive, Baltimore, MD 21218, USA\\
$^{3}$Research School of Astronomy and Astrophysics The Australian National University, Cotter Road Weston Creek, ACT 2611,\\
Australia}

\begin{document}

\date{Accepted 2015 month xx. Received 2015 Month xx; in original form 2015 Month xx}
\pagerange{\pageref{firstpage}--\pageref{lastpage}} \pubyear{2015}

\maketitle

\label{firstpage}

\begin{abstract}
In the previous papers in this series, we have measured the stellar and \hi 
content in a sample of edge-on galaxies.
In the present paper, we perform a simultaneous rotation curve and vertical 
force field gradient decomposition for five of these edge-on galaxies. 
The rotation curve decomposition provides a measure of the radial dark matter 
potential, while the vertical force field gradient provide a measure of the 
vertical dark matter potential.
We fit dark matter halo models to these potentials.
Using our \hi self-absorption results, we find that a typical dark matter 
halo has a less dense core ($0.094\pm0.230$\,M$_\odot$/pc$^3$) compared to 
an optically thin \hi model ($0.150\pm0.124$\,M$_\odot$/pc$^3$).
The \hi self-absorption dark matter halo has a longer scale length $R_c$ 
of $1.42\pm 3.48$\,kpc, versus $1.10\pm 1.81$\,kpc for the optically thin 
\hi model.
The median halo shape is spherical, at $q=1.0\pm0.6$ (self-absorbing \hi), 
while it is prolate at $q=1.5\pm0.6$ for the optically thin.
Our best results were obtained for ESO\,274-G001 and UGC\,7321, for which 
we were able to measure the velocity dispersion in Paper III. 
These two galaxies have drastically different halo shapes, 
with one oblate and one strongly prolate.
Overall, we find that the many assumptions required make this 
type of analysis susceptible to errors.
\end{abstract}


\section{Introduction}\label{sec:dmintro}
Modern day cosmological dark matter simulations predict that dark 
matter has a significant impact on the formation and evolution of galaxies.
The dark matter clumps into haloes, which serve as gravitational 
sinks for baryonic matter to fall in to.
Once inside, these baryons form the galaxies and other visible 
structures in the universe.
The size and shape of these haloes is influenced by the type of 
dark matter particle and its merger history.
As such, getting a grip on the shape of the halo offers a 
potential constraint on the dark matter model \citep{Davis1985A}.

The shape of dark matter haloes can be classified by the shape 
parameter $q$, using the ratios between the vertical axis $c$ 
and radial axis $a$, such that $q = c/a$.
This divides the potential halo shapes up into three classes: 
prolate ($q > 1$), oblate ($q < 1$) and spherical ($q \sim 1$). 
This is, of course, only a simplified version of reality, where 
we can also expect triaxial shapes and changes of shape with radius 
and history \citep{VeraCiro2014}.
However, for haloes with masses $\simgt 10^{12.3} h^{-1} M_\odot$ the 
haloes can be adequately described by a single vertical-to-radial 
axis ratio \citep{Schneider2012A}.

Direct observations of the shape of haloes is tricky, as there are 
only few tracers that offer a clear view on the vertical gravitational 
potential of the halo.
The flat rotation curves of galaxies, while an excellent tracer of the 
radial potential of the halo, provides no information in the vertical direction.
Luckily, some tracers do exist.
The stellar streams of stripped, in-falling galaxies can be used to 
measure the potential the stream is traversing.
\citet{Helmi2004A} analysed the stream of Sagittarius, as found in the 
2MASS survey, around the Galaxy and found that the data best fits a model 
with a prolate shape, with an axis ratio of 5/3.
In a similar fashion, only further out, do the satellites of the Galaxy 
offer such a tracer.
Globular cluster NGC\,5466 was modelled by \citet{Lux2012A} and found 
to favor an oblate or triaxial halo, while excluding spherical and prolate 
haloes with a high confidence.
The stellar stream has been reanalysed by \citet{VeraCiro2013}, who report 
an oblate halo with $q=0.9$ for $r\leq10$\,kpc.
 
Gravitational lensing offers another measure of the vertical gravitational 
potential. 
Strong lensing uses the Einstein lens of background sources around a single 
galaxy or cluster.
By modelling the lens, it is possible to create a detailed mass map of the 
system.
By combining this with gas and stellar kinematics, it is possible to 
calculate the dark matter mass distribution \citep{Treu2010A}.
For example, \citet{Barnabe2012A} applied this method to lensed galaxy 
SDSS\, J2141, to find a slightly oblate halo ($q = 0.91^{+0.15}_{-0.13}$).

Weak lensing lacks the clear gravitation lens seen in strong lensing. 
As such, it is unable to measure the halo of a single galaxy.
Instead, it provides an average halo shape from a statistically large 
sample of sources, by modelling the alignment of background galaxies 
to a large series of foreground galaxies.
This allows the sources to probe the outer edges of haloes.
A recent analysis by \citet{vanUitert2012} finds, on a sample of 
$2.2\times10^7$ galaxies, that the halo ellipticity distribution 
favors oblate, with $q=0.62^{+0.25}_{-0.26}$.

Polar ring galaxies are also of interest in the study of the dark 
matter halo shape, as the orbits of the stars in the polar rings 
are very sensitive to the gravitational potential.
The method was pioneered by \citet{Schweizer1983A} who noted that 
the polar ring in galaxy A0136-0801 indicated a massive halo that 
was more spherical than flat.
\citet{Whitmore1987A} studied this galaxy in more detail and found 
$q=0.98\pm0.20$ for this galaxy.
These authors also studied galaxies NGC\,4650A and ESO\,415-G26, 
for which they report $q=0.86\pm0.21$ and $q=1.05\pm0.17$.
Galaxy NGC\,4650A has also been studied by \citet{Sackett1994A}, 
who reported a flattened halo with $q$ between 0.3 and 0.4.

Another way to place constraints on the halo shape is by carefully 
modelling local edge-on galaxies.
The thickness of the \hi layer in spiral galaxies corresponds directly 
on the local hydrostatic equilibrium.
The \hi layer flares out  at large radii, as there is less matter 
gravitationally binding it to the central plane.
Because of this, flaring provides a sensitive tracer of the vertical
 potential as a function of radius in the disc.
By combining this with the horizontal potential as gained from the 
rotation curve, and estimates for the gas and stellar mass 
distributions, one can fit the potential well created by the dark matter halo.
This method was first applied to the Galaxy by \citet{Celnik1979A} and
by \citet{vdk81c} on edge-on galaxy NGC\,891. In the latter it
was found that the halo was spherical rather than flattened as the stellar disc.
NGC\,4244 was analysed by \citet{Olling1995A}, who found a 
highly flattened halo of $q=0.2^{+0.3}_{-0.1}$.
\citet{OBrien2010D} also set out to  measure the velocity 
dispersion as function of radius.
Using that approach to measure the halo shape of {UGC\,7321}, 
they found a spherical halo ($q = 1.0 \pm 0.1$).

This paper is the fifth in a series. We will refer to earlier papers
of the series simply as Papers I through IV.
In this paper V, we will perform a similar analysis, using the 
measured parameters for the \hi disc (Paper III) 
and stellar disc (Paper IV). The series is based on 
a large part of the PhD thesis
of the first author \citep{Peters2014}.
The paper has the following outline. 
In Section \ref{sec:hydrostaticmodels}, we will provide a detailed 
description of hydrostatic models we use and detail the fitting strategy.
Section \ref{sec:hydrostaticresults} presents and discusses the results.
We summarise the results in Section \ref{sec:hydrostaticssummary}

\section{Modelling Strategy}\label{sec:hydrostaticmodels}
\subsection{Overall Strategy \& Sample}
\citet{Celnik1979A} proposed a new strategy in which the flaring of the \hi 
layers is used as a tracer of the vertical gravitational potential in the 
Galaxy.
This method was extended to edge-on galaxy NGC\,891 by \citet{vdk81c}.
In combination with a traditional rotation curve decomposition, this offers 
a view on both the radial and vertical directions of the gravitational 
potential.
The dark matter halo model can be fit to this.
Further previous work on this subject, using the flaring of gas layers, 
includes that by \citet{Olling1995A,Olling1996A}, \citet{Olling2000},
\citet{bc97}, \citet{nsj05}, \citet{k03}, \citet{kdkh07} and \citet{bj08},
and others.

We will use this same strategy to model the dark matter halo shape 
for the sample of eight galaxies defined in Paper I and extensively studied in
further papers in this series. 
In Paper III, we have measured the structure and kinematics of the \hi in 
these galaxies, using modelling procedures that allow for the 
correction of self-absorption in the \hi for assumed spin 
temperatures. The stellar discs were modelled in Paper IV. 
Based on the quality of the results from Papers III and IV, 
we have decided to model five out of the original eight galaxies in our sample.
These are IC\,5249, ESO\,115-G021, ESO\,138-G014, ESO\,274-G001 and UGC\,7321.
All five galaxies are late-type Sd, with self-absorption corrected \hi masses 
between $4.1\!\pm\!0.1\!\times\!10^8$ and $7.8\!\pm\!0.8\!\times10^9$\,M$_\odot$ 
(Table 1 in Paper III). The distances vary greatly between these galaxies. 
ESO\,274-G001 is only 3.0\,Mpc away, while the distance to IC\,5249 is 
estimated at 32.1\,Mpc.

In Paper III, we successfully measured the velocity dispersion of the 
\hi as function of radius in ESO\,274-G001 and UGC\,7321. 
In ESO\,115-G021, the velocity dispersion appeared to increase with radius
from about 7 \kms in the inner parts to 12-14 \kms at the outer 
measured point.
As we are skeptical of this result, we will adopt a constant velocity 
dispersion of 10\,km/s for this galaxy.
We also adopt this constant velocity dispersion for IC\,5249 and ESO\,138-G014.
For UGC7321 the velocity dispersion drops from 10 \kms in the central 
parts to 8\ \kms at 8 kpc, then remains constant at this level out to 10 kpc, 
after wich it increases to 12\ \kms at 14 kpc. We have sufficient confidence 
in it to adopt it in our analysis.

Galaxy UGC\,7321 has a total B-band magnitude of 13.75 and V-band magnitude 
of 13.19 \citep{Taylor2005}, giving a B-V band difference of 
0.56.\label{sec:estimateML}
To estimate the stellar mass to light ($M_*/L_R$) ratio, we use the model 
interpolation engine\footnote{Available at 
astro.wsu.edu/worthey/dial/dial\_a\_pad.html.}, 
which is based on the \citet{Worthey1994A} and 
\citet{Bertelli1994A} stellar population models.
Using a single burst, Salpeter (slope -2.35) IMF we find 
a good fit using a single population of 950\,Myr and [Fe/H]=0.0, with 
B-V=0.554. 
This would place $M_*/L_R$ in the Rc-band at 0.55.
We have repeated this exercise for our other galaxies, each time finding 
$M_*/L_R\approx0.5$. 
We therefore adopt a mass to light $M_*/L_{R}=0.5$ as a lower boundary to 
the stellar disc mass, and use the reported R-band total luminosities 
reported in Table 3 of Paper IV.
As an upper boundary, we adopt $M_*/L_{R}<3$. At solar metallicity, this 
would imply a single stellar population of approximately 8\,Myr. At 
multiplicities lower than solar, this would increase to an even large age.

\label{sec:MLdiscussion}

\subsection{Decomposition Strategy}
We start with the assumption that there are three components, stars, 
gas+dust and dark matter. Each of these components add to the gravitational 
potential, so ideally one would need to write down and solve the combined 
Poisson-Boltzmann equation. This is internally consistent, but requires
simplifying assumptions such as 
on the properties of the stellar velocity tensor (e.g. that it is Gaussian and
isothermal or a superposition of such components) 
and its variation in space. This approach, which indeed is self-consistent, 
has been used by the earlier workers referred to above, starting with 
\citet{Olling1996A}. 

Our stategy is the following. We infer the radial force near the plane of the 
galaxy from the \hi rotation curve, assuming that due to the relative low 
velocity dispersion of order 10 \kms the asymetric drift term in the Jeans 
equation can be ignored. Next we model the light distribution and assuming a 
constant mass-to-light ratio estimate the contribution of the disk up the a 
free parameter $M/L$. We do the same from the inferred mass distribution of 
gas, adding the usual contribution (25\%\ of the total) for helium. We assume 
also that the molecular gas content in these late-type dwarfs is small. This 
gives us (except for the parameter $M/L$) the gradient in the radial force 
due to the halo. We next estimate the vertical force from the thickness and 
velocity dispersion of the \hi, using the vertical Jeans equation. Since we 
evaluate the forces at $z = 100$ pc, this  should be 
an excellent approximation, even though the equations used are not internally
self-consistent as in the approach in the first paragraph.
This strategy has previously been used by \citet{OBrien2010D}.

\subsubsection{Radial Tracer}
The radial force gradient is calculated using a classic rotation curve 
decomposition performed at the mid-plane of the galaxy \citep{vanAlbada1985A}:
\begin{equation}
 v_\textrm{total}^2(R) = v_\textrm{gas}^2(R) + v_\textrm{stellar}^2(R) + v_\textrm{halo}^2(R)\,\,.\label{eqn:vresidual}
\end{equation}
The total rotation $v_\textrm{total}(R)$ is the observed rotation curve of 
the \hi gas, as we measured previously in Paper III. 
The theoretical rotation curves of the gas $v_\textrm{gas}(R)$ and stars 
$v_\textrm{stellar}(R)$  represent the contributions due to the stellar 
and gaseous mass components.
We calculate the theoretical rotation curve of the stellar and gaseous 
components, using equation (A.17) of \citet{Casertano83}
\begin{equation}
v^2(R) = -8GR \int_0^\infty r \int_0^\infty \frac{\partial \rho(r,z)}{\partial r} \frac{K(p)-E(p)}{(Rrp)^{1/2}} \,dz\,dr \,\,,
\end{equation}
with 
\begin{eqnarray}
p &=& x - (x^2 -1)^{1/2}\,\,,\\
x &=& \frac{R^2 + r^2 + z^2}{2Rr}\,\,.
\end{eqnarray}
Here $K(p)$ and $E(p)$ are the complete elliptical integrals of the first 
and second kind. 
The equations are evaluated numerically.
The equation of the density distribution $\rho(r,z)$ of the \hi disc has 
been previously presented in Equation 15 in Paper II. 
We have presented the measured densities in Paper III.
The equations for the stellar disc and bulge luminosity distributions have 
been previously presented in Equations 1 and 3 in Paper IV.
We convert to a combined stellar mass distribution from these equations using
\begin{equation}
 \rho(r,z) = (M_*/L_R)\left(j_\textrm{disc}(R,z) + j_\textrm{bulge}(R,z)\right)\,\,.
\end{equation}
The measurements for the stellar disc have been presented in Paper IV.
Note that we adopt a single mass-to-light ratio $M_*/L_R$.
While this may be a bit unrealistic, we noted in Paper IV that the fits for 
some of the bulges appear to supplement the stellar disc, rather than 
model a central component.
Treating the two as distinct would therefore be invalid.
A fixed $M_*/L_R$ ratio also has the advantage of limiting the complexity 
of the parameter space we will be fitting in.

\subsubsection{Vertical Tracer}
We follow the method of \citet{OBrien2010D} to calculate the gradient of the 
vertical force.
The gradients due to each mass component add up as
\begin{eqnarray}
\frac{dF_{z,\textrm{total}}(R,z)}{dz} &=& \nonumber \\
\frac{dF_{z,\textrm{gas}}(R,z)}{dz}&+&\frac{dF_{z,\textrm{stellar}}(R,z)}{dz}+\frac{dF_{z,\textrm{halo}}(R,z)}{dz} \label{eqn:zresidual}
\end{eqnarray}

Following \citet{OBrien2010D}, the disc is assumed to be in vertical 
hydrostatic equilibrium, such that the vertical gas pressure gradient 
and total vertical gravitational force of the galaxy potential 
$\Phi_\textrm{total}$ due to all mass components balance perfectly, 
\begin{equation}
\nabla\left(\sigma_\textrm{gas}^2 \,\rho_\textrm{gas}\right) = \rho_\textrm{gas} \nabla \Phi_\textrm{total}\label{eqn:hydro}\,\,.
\end{equation}

We next assume that the gas velocity dispersion is isothermal\footnote{Note 
that we have also discussed this assumption in Section 7 of Paper IV.} in $z$.
In that case, Equation \ref{eqn:hydro} reduces to
\begin{equation}
\frac{\sigma_\textrm{gas}^2 \,\partial\left[ \log \rho_\textrm{gas}(R,z)\right]}{\partial z^2} = - \frac{\partial F_{z,\textrm{total}} (R,z)}{\partial z}\,\,.\label{assumption:isothermal}
\end{equation}

Since our \hi disc is modelled as a Gaussian distribution (see Equation 
15 of Paper II), this becomes
\begin{equation}
\frac{\partial F_{z,\textrm{total}} (R,z)}{\partial z} = -\frac{\sigma^2_\textrm{gas}(R)}{{z_0}^2(R)}\label{eqn:Ftotal}\,\, ,
\end{equation}
such that the vertical gradient of $F_\textrm{z,total}$ is constant with 
height $z$.

The gradients of the stellar and gas force components was calculated 
using the Poisson equation, where we assume the disc is axi-symmetric 
and the circular rotation is constant with height \citep{OBrien2010D},
\begin{equation}
\frac{\partial F_{z} (R,z)}{\partial z} = -4\pi G \rho(R,z) + \frac{1}{R} \frac{\partial \, v^2(R)}{\partial R},\label{eqn:poissonz}
\end{equation}
where we use the density $\rho$ and rotation $v$ of each of the two components.
The squared velocity gradient is calculated numerically.
Our modelling of the vertical tracer will use a plane at a height $z$ 
of 100\,pc.

\subsection{Halo Potential}
We make use of the flattened, but axi-symmetric, pseudo-isothermal halo model 
proposed by \citet{Sackett1994A}, where we assume that the equatorial plane 
of the halo matches that of the galaxy.
In this model the density is stratified in concentric ellipsoids, as given by 
\begin{equation}
\rho_\textrm{halo}(R,z) = \frac{\rho_{0,\textrm{halo}}\,R_c^2}{R_c^2 + R^2 + z^2/q^2}.
\end{equation}
The ellipsoids formed by this density distribution have axis ratio 
$q\equiv c/a$, with core radius $R_c$.

The potential due to this density distribution is given by 
\citet{Sackett1990A} as
\begin{eqnarray}
\Phi_\textrm{halo}(R,z) &=\ \  2 \pi G q \,\rho_{0,\textrm{halo}} \,R_c^2 \int_0^{1/q} \left[\frac{1}{x^2(1-q^2) + 1}\right]&\nonumber \\ &\times  \log\left\{1+\frac{x^2}{R_c^2} \left[\frac{R^2}{x^2(1-q^2) +1}+z^2\right]\right\}\,dx.& 
\end{eqnarray}

\citet{Sackett1994A} provides the solution to the forces associated with this 
halo in spherical coordinates:
\begin{eqnarray}
F_R(R,z) &=& \frac{-v_H^2 R \gamma}{h \arctan \gamma} \left[ \frac{\mu^2}{\mu^2-1} \left(\frac{\arctan \gamma\mu}{\gamma\mu} - \frac{\arctan\gamma}{\gamma}\right)\right.\label{eqn:halor} \nonumber \\ 
&&\left.- \frac{\nu^2}{\nu^2-1} \left(\frac{\arctan \gamma\nu}{\gamma \nu} - \frac{\arctan \gamma}{\gamma} \right) \right] \,,\\
F_z(R,z) &=& \frac{-v_H^2 z \gamma}{h \arctan \gamma}\left(\frac{\arctan \gamma\mu}{\gamma \mu} - \frac{\arctan \gamma \nu}{\gamma \nu}\right)\,,\label{eqn:haloz}
\end{eqnarray}
where
\begin{equation}
\gamma \equiv \frac{\sqrt{1-q^2}}{q}\,\, ,\,\, \nu\equiv \sqrt{\frac{2c}{b-h}}\,\,,\,\, \mu\equiv\sqrt{\frac{2c}{b+h}},
\end{equation}
and
\begin{eqnarray}
h&\equiv&\sqrt{b^2-4ac}  \, ,\\
a &\equiv& (1-q^2) R_c^2 \, ,\\
b &\equiv& z^2 + R^2 + (1-q^2) R_c^2 \, ,\\
c &\equiv& z^2.
\end{eqnarray}
Note that to calculate $q>1$ one requires the use of complex numbers, although these return to real numbers at the end of the calculation. Only using real numbers will force a fit to be constrained to $q<1$.
The calculation has a singularity at $q=1$, which we automatically substitute with $q=0.999$ where required.

The asymptotic halo velocity $v_H$ is defined as 
\begin{equation}
v_H^2 = \frac{4\pi G\,\rho_{0,\textrm{halo}} \,R_c^2\, q \arccos q}{\sqrt{1-q^2}}\label{eqn:dmhalovelocity}
\end{equation}

\subsection{Fitting Strategy}

Using Equation \ref{eqn:vresidual}, we calculate the observed circular 
rotation curve due to the dark matter halo $v_\textrm{halo}^2$ at the 
mid-plane of the galaxy over the full range of $R$.
In a similar way we calculate the observed vertical force gradient  
$dF_{z,\textrm{halo}}/dz(R,z)$ due to the dark matter halo at a height $z$ 
of 100\,pc, using Equation \ref{eqn:zresidual}.
We inspect the results to determine in which range of radii $R$ they 
are sufficiently reliable.
In contrast to \citet{OBrien2010D} we fit both the vertical and radial 
tracers simultaneously.
Since the rotation curve decomposition is less sensitive to noise than 
the vertical force gradient decomposition, we can often fit a larger 
radial range for the rotation curve decomposition  than the vertical 
force gradient decomposition.
We will fit the dark matter halo using Equations \ref{eqn:halor} and 
\ref{eqn:haloz}, where we calculate the gradients numerically, and use 
the mid-plane approximation $v^2=-RF$ \citep{Kuijken1989A}.

Both tracers operate in drastically different numerical regimes. 
The total observed rotation curve is often near 100\,km/s, while the 
observed vertical force gradient has a value of -0.004 km$^2$ s$^{-2}$ pc$^{-2}$.
As these numbers are so drastically different, the combined $\chi^2$ 
would be dominated by the rotation curve decomposition.
We therefore  normalise the data of each force by its maximum value 
in that range, such that the total $\chi^2$ error is calculated as
\begin{eqnarray}
\chi^2 &=& \chi^2_R + \chi^2_z\,\label{eqn:halochi}\, ,\\
\chi^2_R &=& \sum_{R_i} \frac{v_\textrm{total}^2 - v_\hi^2 - v_\textrm{stellar}^2 - v_\textrm{halo}^2}{\max({v_\textrm{total})^2}}\,\,,\label{eqn:halochi1}\\
\chi^2_z &=& \sum_{R_i} \frac{\left(  dF_{z,\textrm{diff}}/dz   \right)^2}{\min(dF_{z,\textrm{total}}/dz)^2}\,\,,\label{eqn:halochi2}\\
\frac{dF_{z,\textrm{diff}}}{dz} &=& \frac{dF_{z,\textrm{total}}}{dz} \nonumber \\
&-& \frac{dF_{z,\textrm{gas}}}{dz}+\frac{dF_{z,\textrm{stellar}}}{dz}+\frac{dF_{z,\textrm{halo}}}{dz}\,\,,
\end{eqnarray}
where the $R_i$ values are all data points within the respective 
fitting ranges used for the vertical and radial directions.
We have tested many variations of Equations \ref{eqn:halochi1} and 
\ref{eqn:halochi2}, including converting and integrating both the 
tracers back into forces so that they could compared more directly. 
The overall problem however remained, as the two forces were too 
different in strength and the errors in the radial component would 
dominate the fit.
We have decided to stick to the units adopted by \citet{OBrien2010D}, 
as this offered us the best way to compare results.

In Paper III, we performed a Monte-Carlo Markov-Chain (MCMC) fit to the 
neutral hydrogen, such that a set of samples from the so-called chain 
together cover the likelihood distribution of the parameters.
Our fitting strategy here makes use of this likelihood distribution. 
We take the last 1000 samples from the chain and perform a fit of the 
dark matter halo to each individual sample.
In total, we thus get 1000 solutions for the halo.
We make use of the \textsc{PSwarm} particle swarm optimization 
algorithm \citep{Vaz09}, as implemented through the \textsc{OpenOpt} 
library \citep{Kroshko}.
The \textsc{PSwarm} is an example of a global optimization routine, 
which can avoid being stuck in local optima in the solution space.
The fit is performed directly to $\chi^2$ as defined in Equation 
\ref{eqn:halochi}.
In some cases, the models converge to unrealistic solutions.
We therefore base our results on the 25\% of the samples with the 
lowest $\chi^2$ errors. 
The halo model uses three free parameters: the halo central density 
$\rho_{0,\textrm{halo}}$, the scale length $R_c$ and the halo shape $q$.
Together with the mass-to-light conversion $M_*/L_{R}$ for the stellar 
disc, we thus have  four free parameters.
We have considered using an additional mass-to-light conversion for 
the bulge, but found that with these four parameters the solutions 
were already becoming degenerate. Adding an additional free parameter 
would have worsened this problem. 

We constrain  $\rho_{0,\textrm{halo}}$ between 0 and 3 M$_\odot$ pc$^{-3}$, $R_c$ 
between 
100\,pc and 10\,kpc, $q$ between 0.1 and 2.0, and $M_*/L_{R}$ between 0.5 
and 3.0.

\begin{figure*}
\centering
\includegraphics[width=0.40\textwidth]{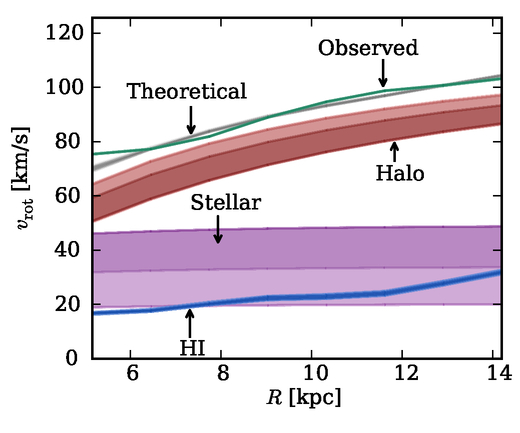}
\includegraphics[width=0.40\textwidth]{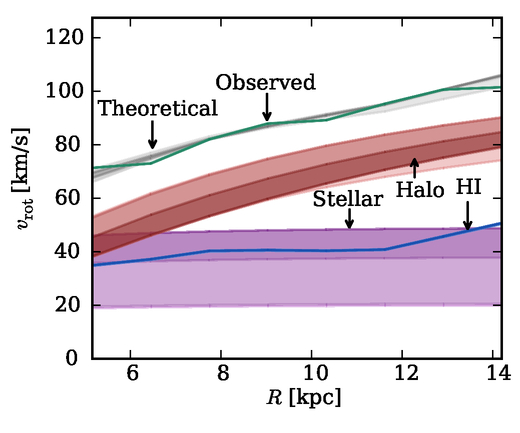}
\includegraphics[width=0.40\textwidth]{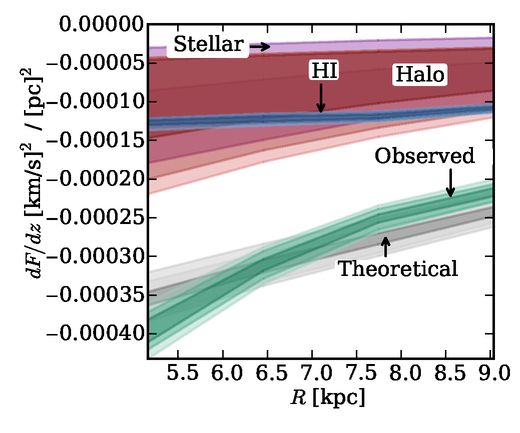}
\includegraphics[width=0.40\textwidth]{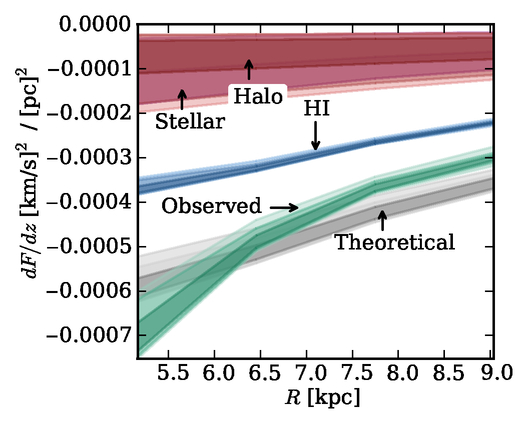}
\caption[Halo decomposition for IC\,5249]{Rotation curve and vertical force gradient decomposition of IC\,5249. The left side panels show the results for the optically thin \hi models, while the right side panels show the results for the self-absorption at $T_\textrm{spin}=100$\,K results.}\label{fig:IC5249decomposition}
\end{figure*}

\begin{figure*}
\centering
\includegraphics[width=0.49\textwidth]{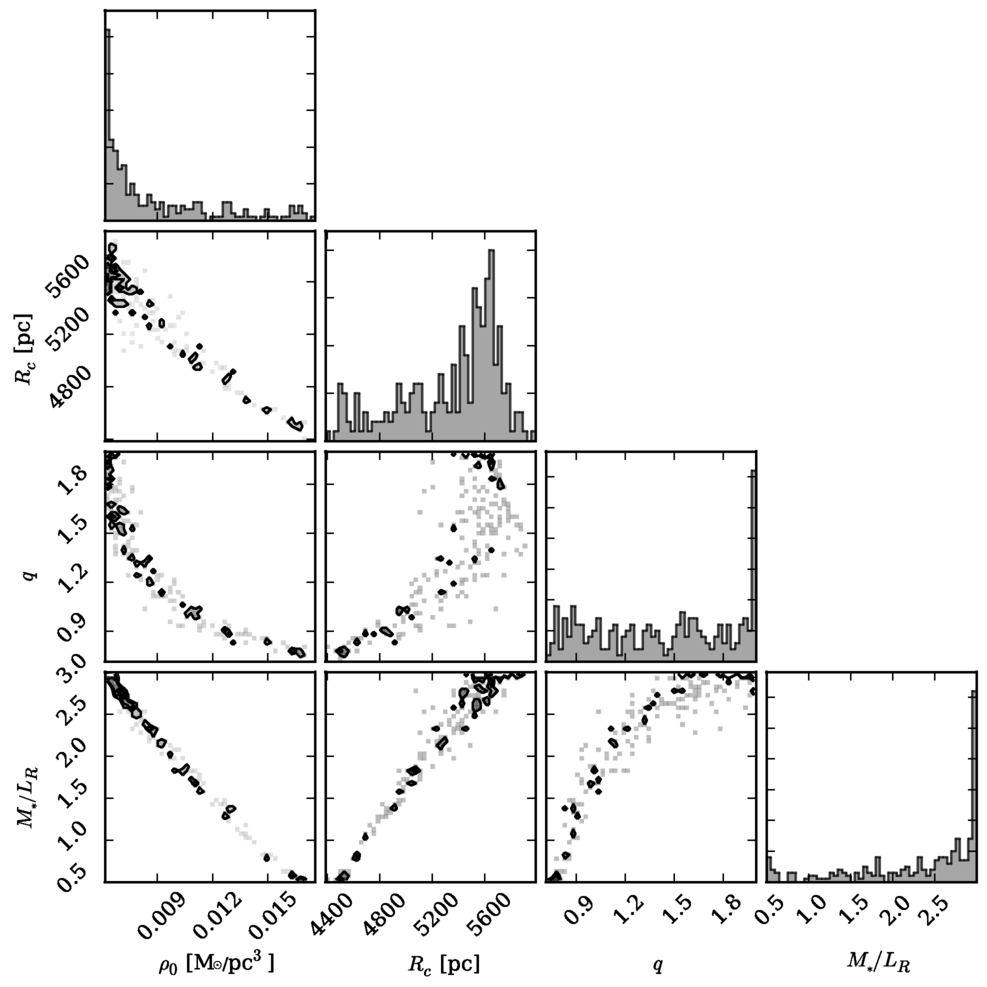}
\includegraphics[width=0.49\textwidth]{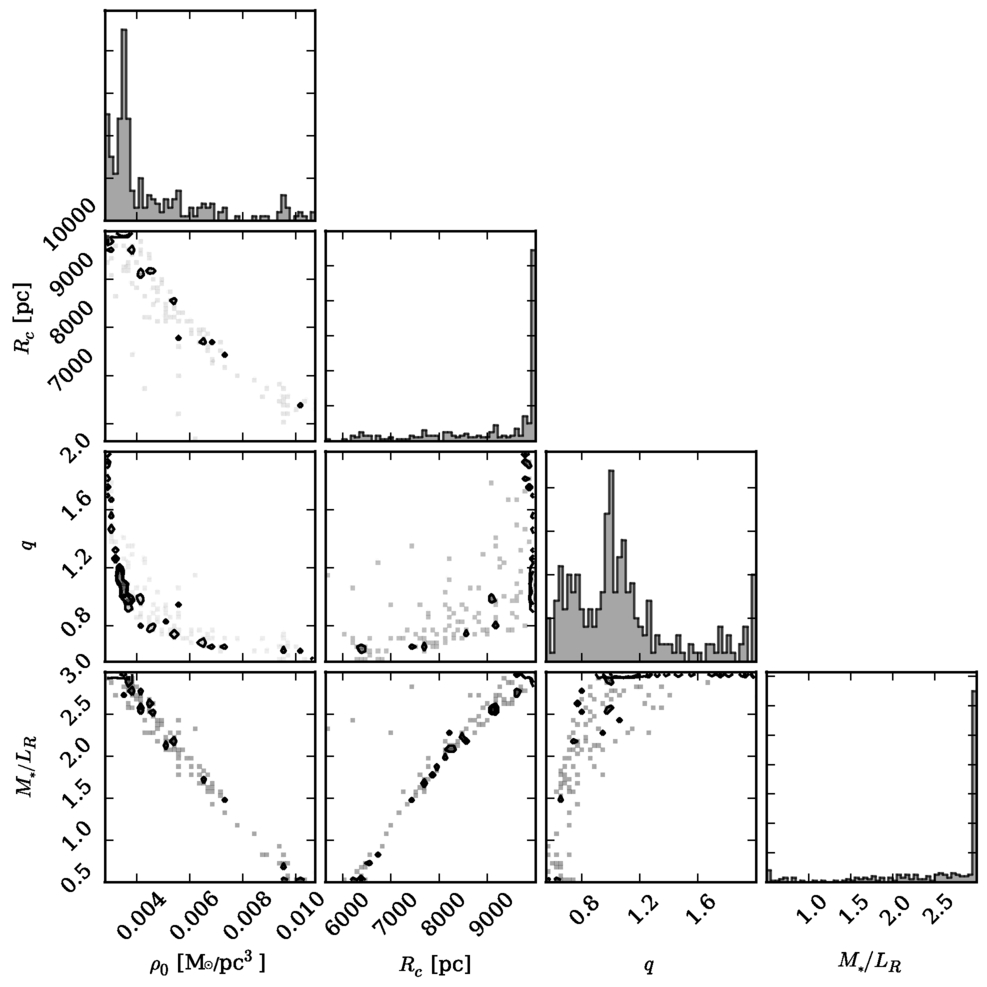}
\caption[Halo parameters of IC\,5249]{Correlation diagram for the optically thin model of IC\,5249 (left) and the self-absorption model (right), both with a constraint of M/L$\geq0.5$.}\label{fig:IC5249-crosscorr}
\end{figure*}

\begin{figure*}
\centering
\includegraphics[width=0.40\textwidth]{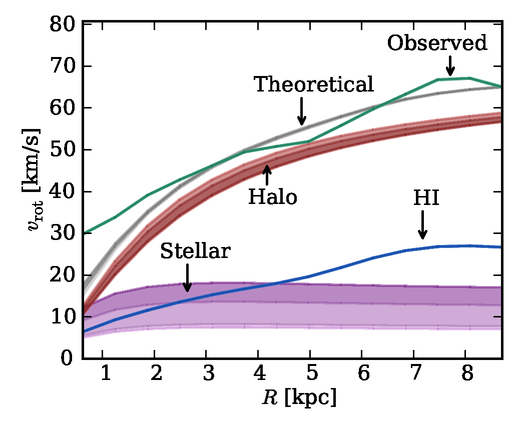}
\includegraphics[width=0.40\textwidth]{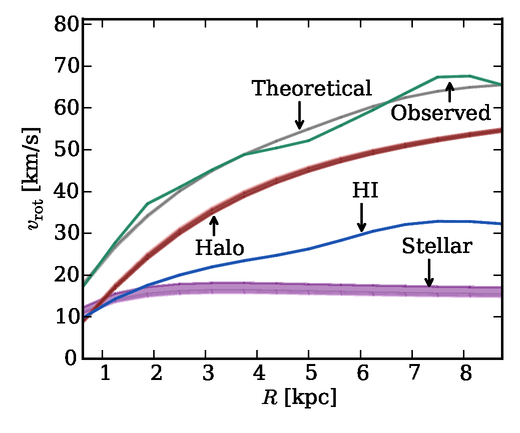}
\includegraphics[width=0.40\textwidth]{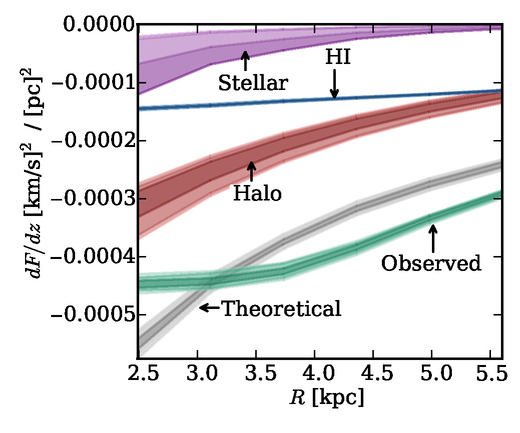}
\includegraphics[width=0.40\textwidth]{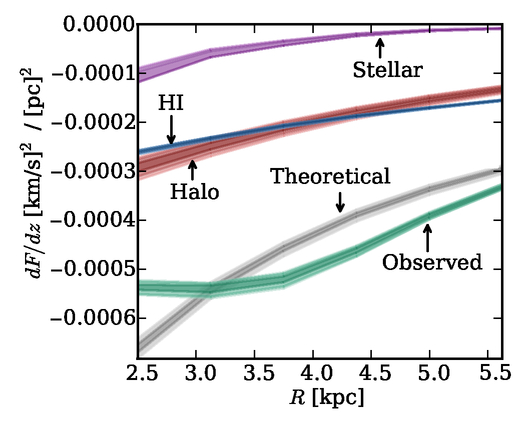}
\caption[Halo decomposition for ESO\,115-G021]{Rotation curve and vertical force gradient decomposition of ESO\,115-G021. The left side panels show the results for the optically thin \hi models, while the right side panels show the results for the self-absorption at $T_\textrm{spin}=100$\,K results.}\label{fig:ESO115decomposition}
\end{figure*}

\begin{figure*}
\centering
\includegraphics[width=0.49\textwidth]{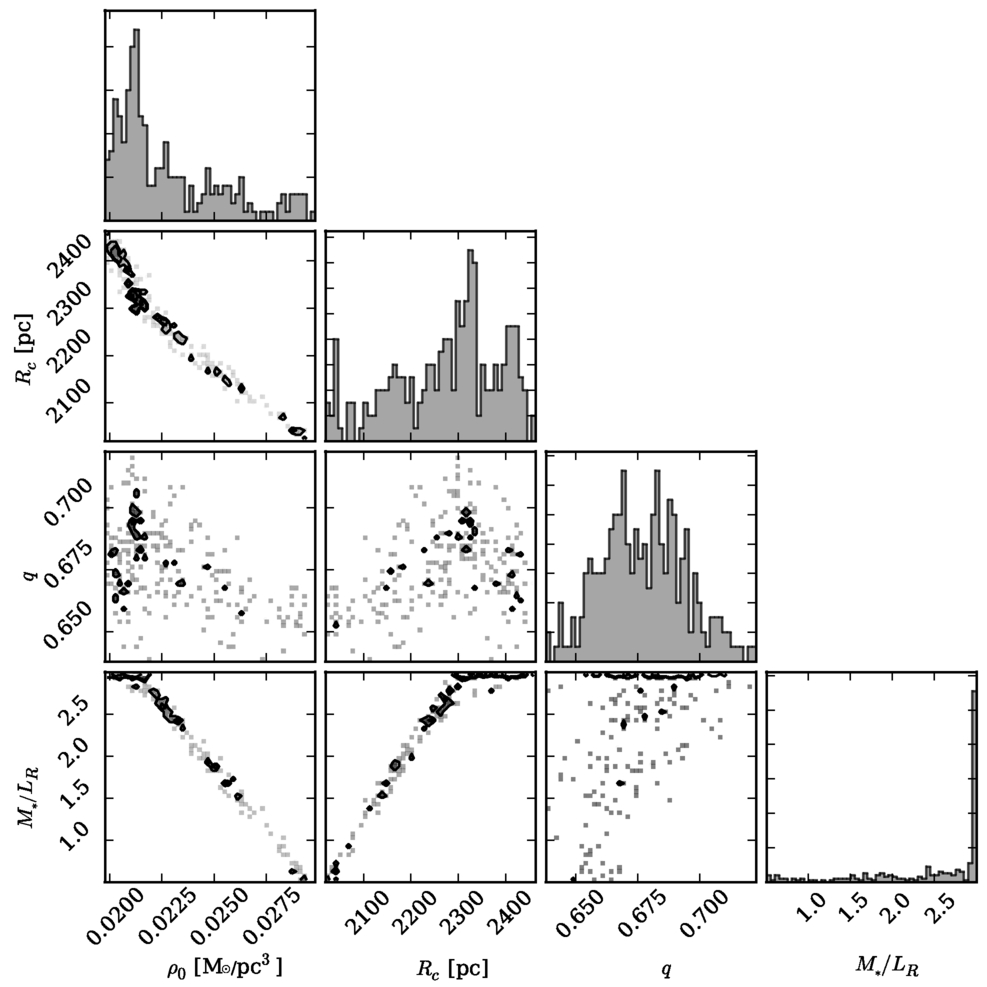}
\includegraphics[width=0.49\textwidth]{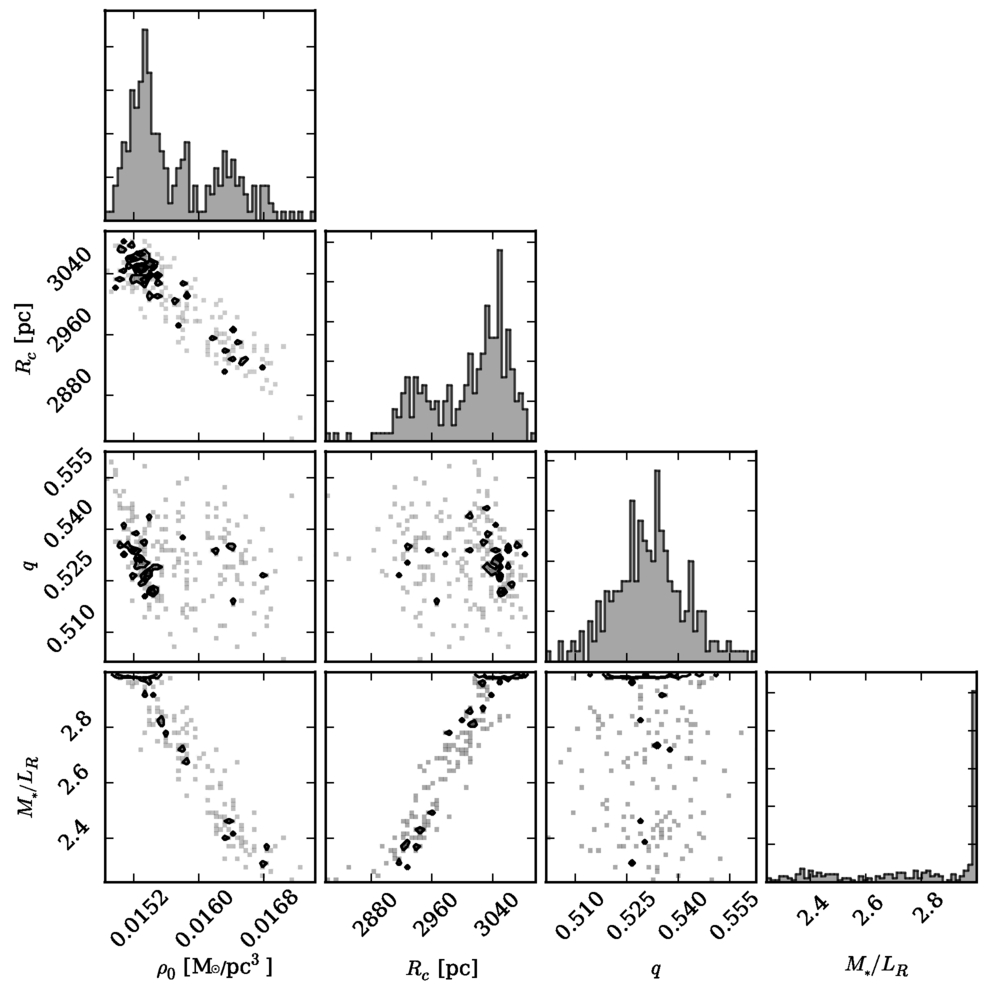}
\caption[Halo parameters of ESO\,115-G021]{Correlation diagram for the optically thin model of ESO\,115-G021 (left) and the self-absorption model (right), both with a constraint of M/L$\geq0.5$.}\label{fig:ESO115-crosscorr}
\end{figure*}

\begin{figure*}
\centering
\includegraphics[width=0.40\textwidth]{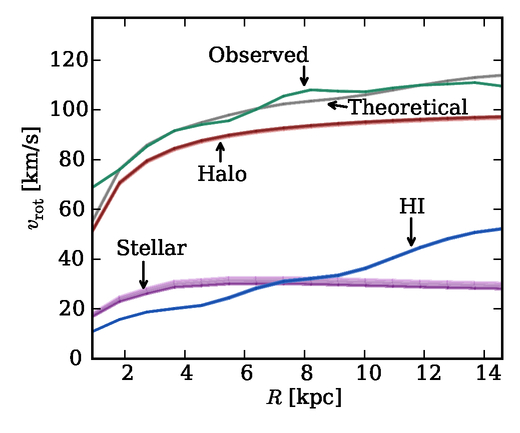}
\includegraphics[width=0.40\textwidth]{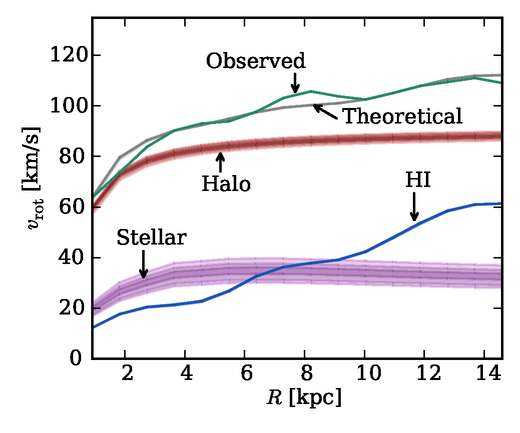}
\includegraphics[width=0.40\textwidth]{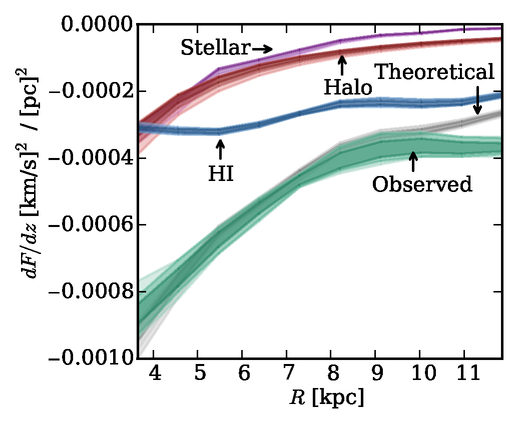}
\includegraphics[width=0.40\textwidth]{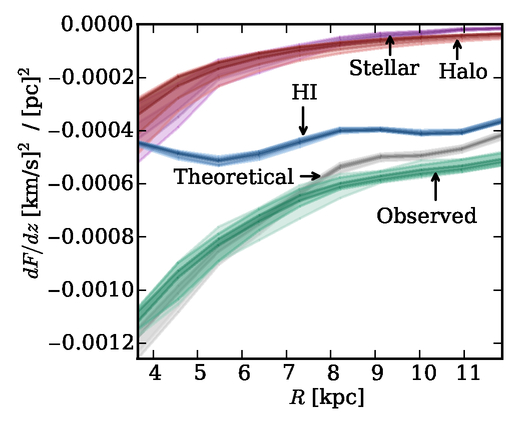}
\caption[Halo decomposition for ESO\,138-G014]{Rotation curve and vertical force gradient decomposition of ESO\,138-G014. The left side panels show the results for the optically thin \hi models, while the right side panels show the results for the self-absorption at $T_\textrm{spin}=100$\,K results.}\label{fig:ESO138decomposition}
\end{figure*}

\begin{figure*}
\centering
\includegraphics[width=0.49\textwidth]{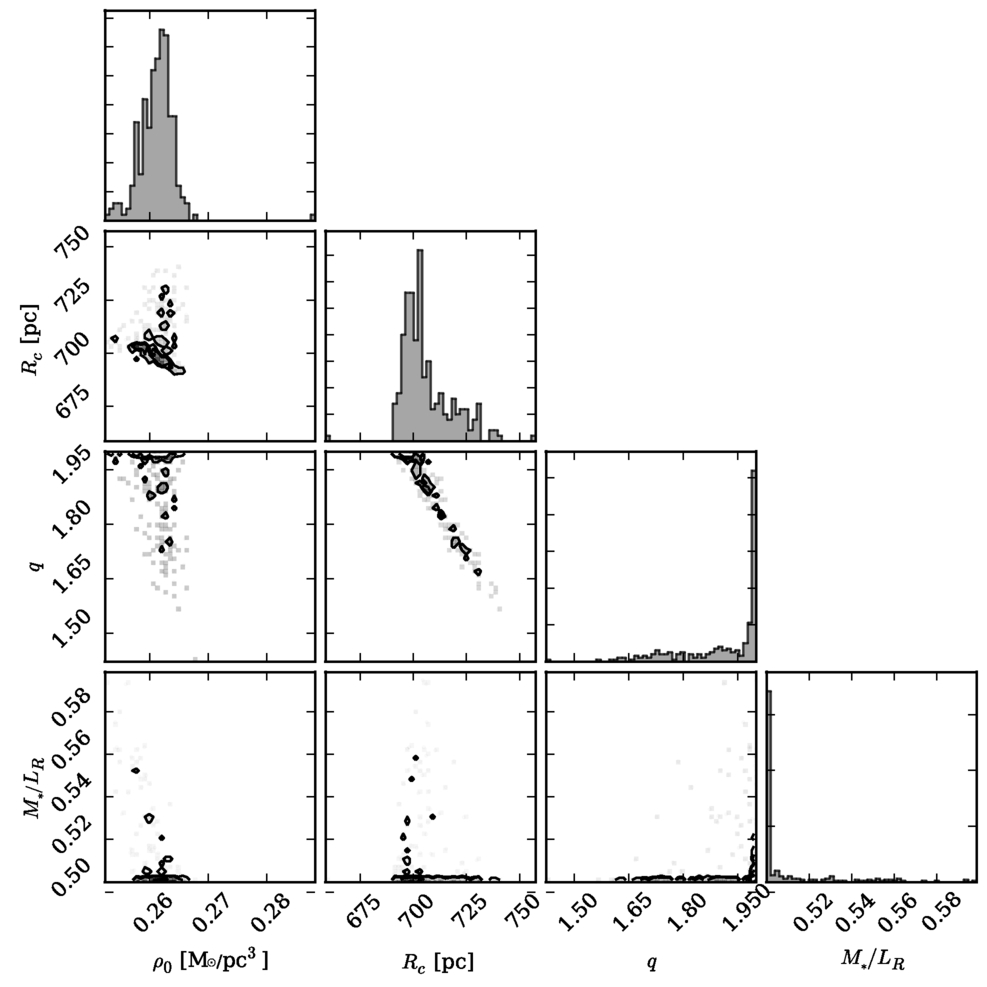}
\includegraphics[width=0.49\textwidth]{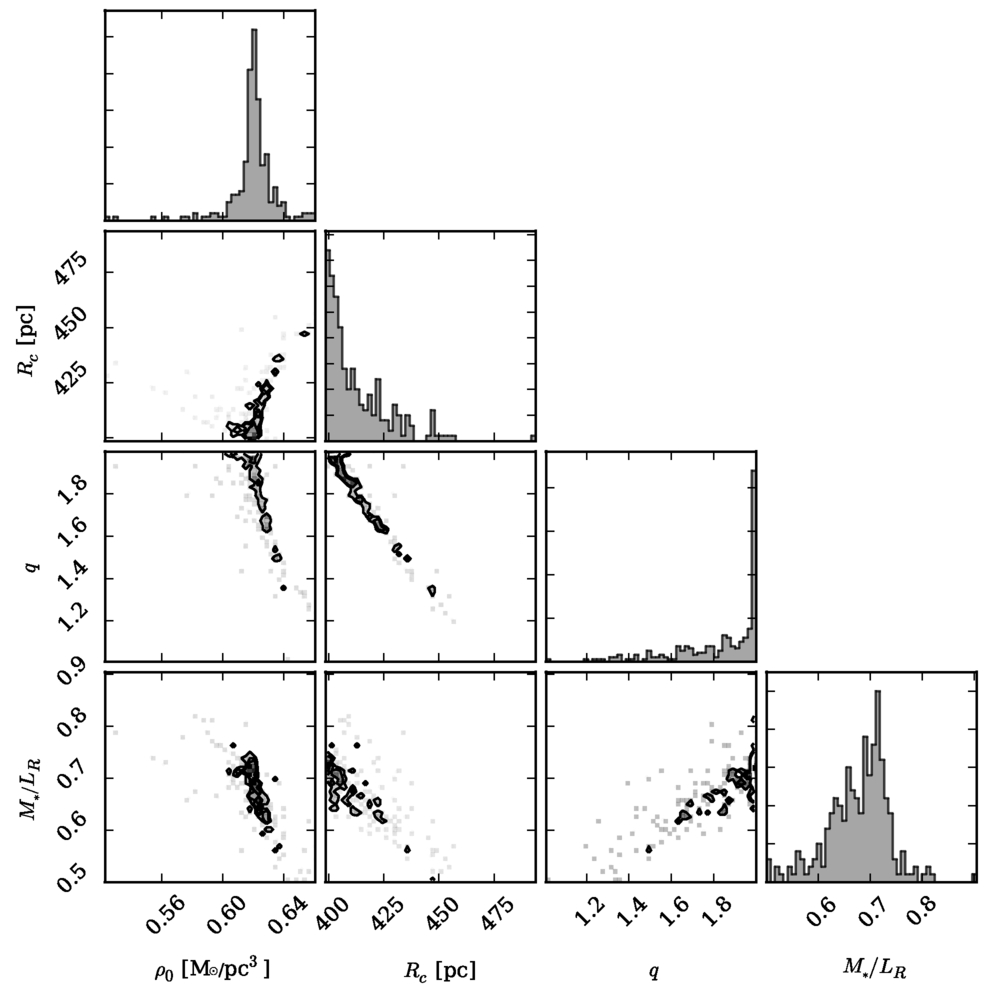}
\caption[Halo parameters of ESO\,138-G014]{Correlation diagram for the optically thin model of ESO\,138-G014 (left) and the self-absorption model (right), both with a constraint of M/L$\geq0.5$.}\label{fig:ESO138-crosscorr}
\end{figure*}

\begin{figure*}
\centering
\includegraphics[width=0.40\textwidth]{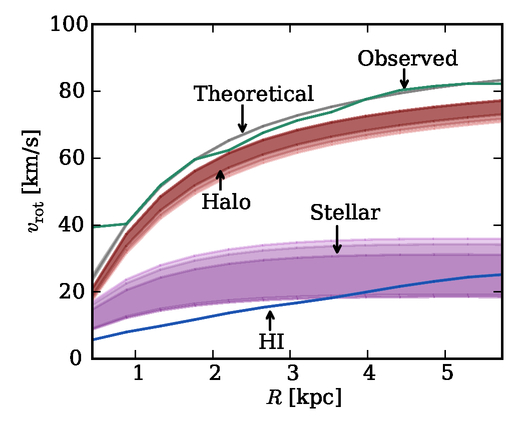}
\includegraphics[width=0.40\textwidth]{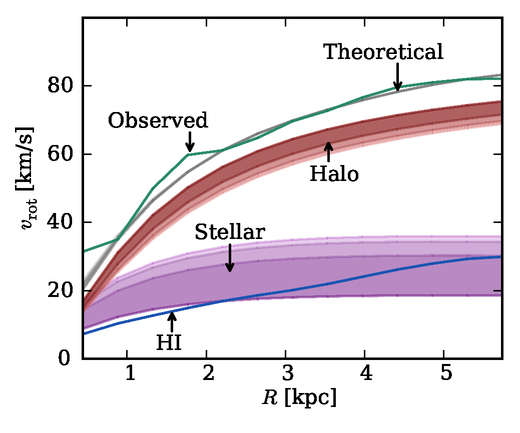}
\includegraphics[width=0.40\textwidth]{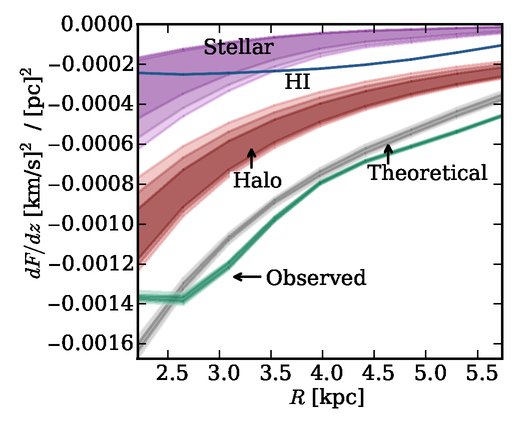}
\includegraphics[width=0.40\textwidth]{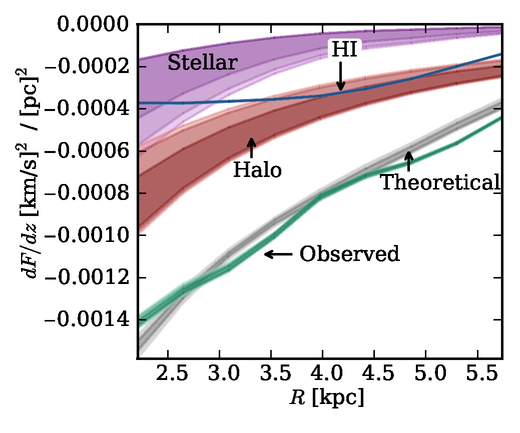}
\caption[Halo decomposition for ESO\,274-G001]{Rotation curve and vertical force gradient decomposition of ESO\,274-G001. The left side panels show the results for the optically thin \hi models, while the right side panels show the results for the self-absorption at $T_\textrm{spin}=100$\,K results.}\label{fig:ESO274decomposition}
\end{figure*}

\begin{figure*}
\centering
\includegraphics[width=0.49\textwidth]{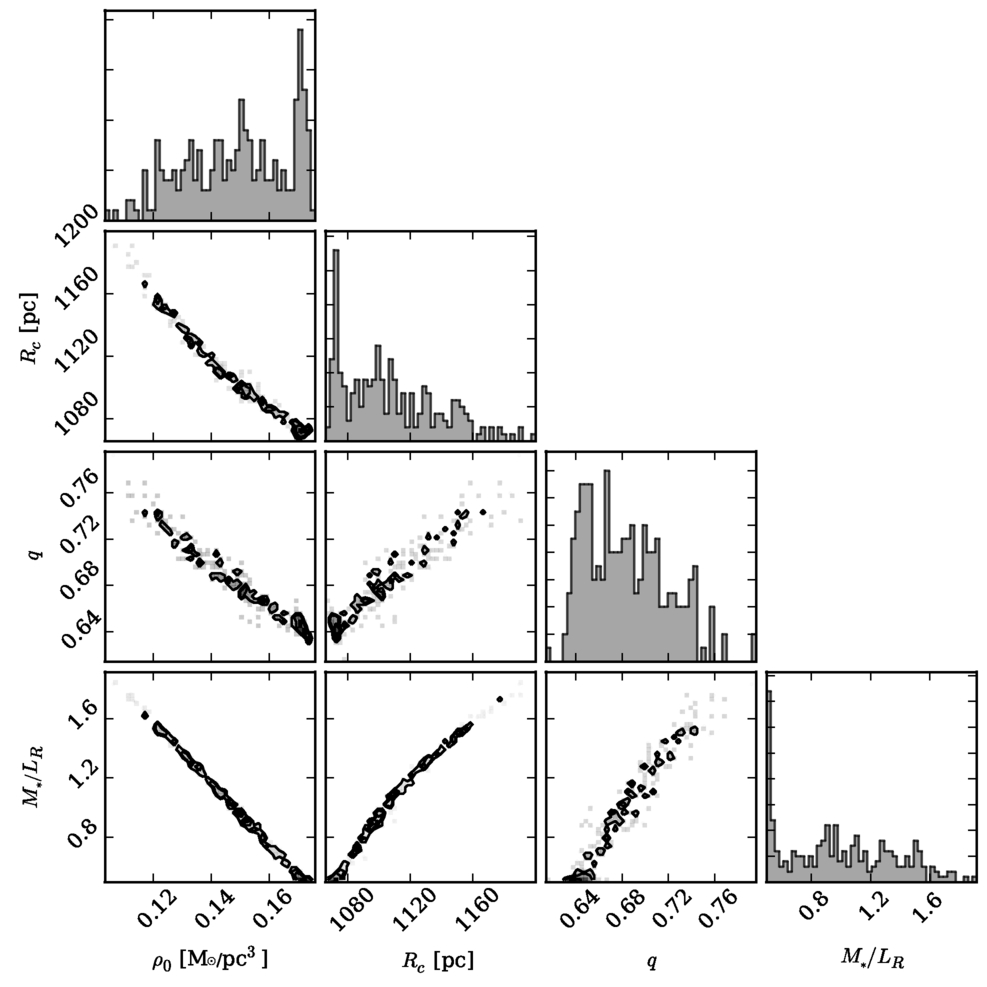}
\includegraphics[width=0.49\textwidth]{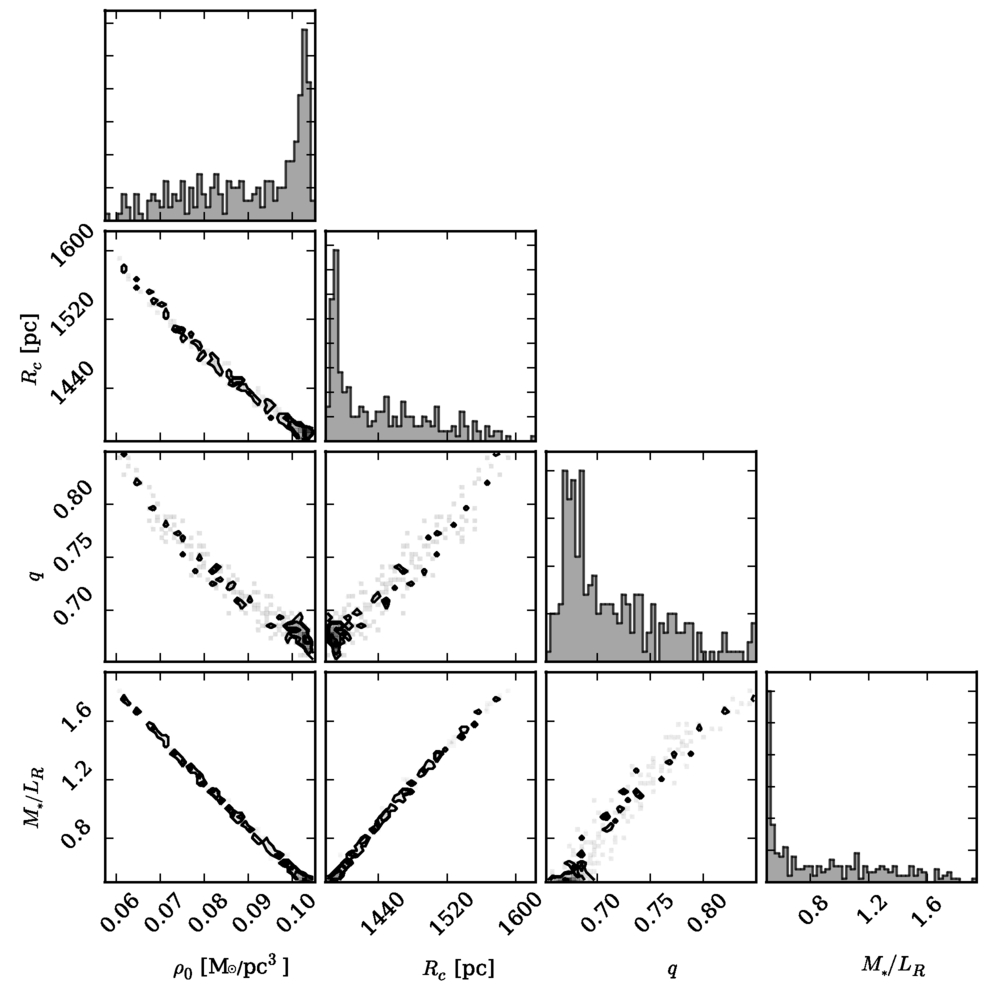}
\caption[Halo parameters of ESO\,274-G001]{Correlation diagram for the optically thin model of ESO\,274-G001 (left) and the self-absorption model (right), both with a constraint of M/L$\geq0.5$.}\label{fig:ESO274-crosscorr}
\end{figure*}

\section{Results}\label{sec:hydrostaticresults}
\subsection{IC\,5249}
We have been moderately successful in modelling the dark matter halo of 
galaxy IC\,5249. 
The decomposition of both the optically thin and self-absorbing \hi 
results is shown in Figure \ref{fig:IC5249decomposition}.
As is clear from the figure, the uncertainties in the stellar halo 
contribution, and subsequently the dark matter halo contribution are 
quite severe.
This is mostly due to the fact that the measurements of both the vertical 
force gradient, as well as the rotation curve, start relatively far out 
(near 5.0-5.5\,kpc). 
The data of the  inward parts of the galaxy were too uncertain for a 
reliable measurement of the tracers.
As the dark matter halo shape $q$ can most accurately be constrained 
from the vertical force gradient in the inner parts of the galaxy 
\citep[see][]{OBrien2010D}, this lack of data does not allow us to 
constrain $q$ significantly.
The optically thin \hi model yields $q=1.5^{+0.5}_{-0.5}$, while the 
self-absorbing \hi yields $q=1.0^{+0.4}_{-0.3}$. 

The lack of a significant constraint on $q$ leads to strong correlations 
between the other parameters. 
This is reflected in the cross-correlation diagrams for the parameters, 
seen in Figure \ref{fig:IC5249-crosscorr} (left) for the optically thin case, 
and Figure \ref{fig:IC5249-crosscorr} (right) for the self-absorbing \hi case.
An oblate dark matter halo shape ($q<1$) produces a less massive stellar 
disc, with a shorter dark matter halo scale length $R_c$ and higher dark 
matter halo core density $\rho_0$. 
This behavior holds in both models. 
Over the whole dataset, we find that for an optically thin \hi model the 
halo is found at 
$\rho_0 =0.007^{+0.005}_{-0.001}$\,M$_\odot$/pc$^3$, $R_c=5.45^{+0.21}_{-0.52}$\,kpc. 
The stellar disc is found with $M_*/L_R=2.62^{+0.38}_{-1.19}$.
The self-absorption \hi models return 
$\rho_0 = 0.004^{+0.003}_{-0.001}$\,M$_\odot$/pc$^3$, 
a scale length $R_c$ of $9.79^{+0.21}_{-2.06}$, and 
a stellar disc with $M_*/L_R=2.98^{+0.02}_{-1.17}$.

As we discussed in Section \ref{sec:MLdiscussion}, the most likely 
$M_*/L_R$ values lie close to 0.5. If we thus limit ourselves to the 
data points at $M_*/L_R<0.55$, we find for the optically thin \hi an 
oblate halo with
$q =0.76_{-0.03}^{+0.04}$.
The core density of the dark matter halo is 
$\rho_0 = 0.017_{-0.001}^{+0.001}$\,M$_\odot$/pc$^3$ and its scale length is
$R_c = 4.49_{-0.05}^{+0.02}$\,kpc.
The self-absorbing \hi model returns an even more oblate halo, with a shape of 
$q =0.55_{-0.03}^{+0.14}$, a core density of 
$\rho_0 = 0.014_{-0.003}^{+0.003}$\,M$_\odot$/pc$^3$ and a scale length of 
$h_0 = 5.19_{-1.01}^{+1.30}$\,kpc.
Comparing the two models, we see that the dark matter halo of the 
self-absorbing \hi requires a more oblate halo, with longer scale 
length $R_c$ and less massive central density $\rho_0$.

\subsection{ESO\,115-G021}
We have not been very successful in modelling ESO\,115-G021.
As can be seen from the results in Figure \ref{fig:ESO115decomposition}, the 
observed vertical force gradient has a nearly flat slope near the inner parts 
of the galaxy ($R<4$\,kpc). 
This is problematic to fit to, as we always expect the vertical force gradient 
to get increasingly strong near the inner parts. 
We have attempted fitting only beyond $R=3.5$\,kpc, but this left only 2\,kpc 
in which we could fit the data, which did not result in a stable fit.
Smoothing was applied on the input parameters, but this did not improve the 
quality of the observed vertical force gradient.
As such, we hereby present our best fit, but encourage the reader to have a 
skeptical treatment of them.

The self-absorbing \hi model of the galaxy results in a dark matter core 
density $\rho_0$ of $0.015^{+0.001}_{-0.001}$\,M$_\odot$/pc$^3$, a scale length 
$R_c$ of $3.02^{+0.03}_{-0.08}$\,kpc and an oblate shape of $q=0.5^{+0.1}_{-0.1}$.
 The stellar disk is found to have a high $M_*/L_R=2.89^{+0.11}_{-0.44}$.
The optically thin \hi model produces a more massive central core density 
$\rho_0=0.022^{+0.004}_{-0.001}$\,M$_\odot$/pc$^3$, a shorter scale length of 
$R_c=2.30^{+0.10}_{-0.15}$ and a less oblate halo shape  $q=0.7^{+0.1}_{-0.1}$. 
We again find a high $M_*/L_R=2.84^{+0.16}_{-1.15}$\\
The cross-correlation diagrams of the models are shown in 
Figure \ref{fig:ESO115-crosscorr}.

\subsection{ESO\,138-G014}\label{sec:DM138}

Galaxy ESO\,138-G014 was initially hard to model, as the total observed 
vertical force gradient was already weaker than the contribution from  
the neutral hydrogen alone.
As we noted before in Paper III, the galaxy seems to 
have quite a thick \hi layer (see Figure 13 of Paper III). 
The most likely explanation for this is that the galaxy is not seen 
completely edge-on.
This is consistent with the observed stellar disc from Paper IV, 
where we measured $i=86.8^\circ$ (Table 2 of Paper IV).
We have attempted to correct for this by lowering the observed thickness 
by 30\%, but will treat the results are uncertain in view of this
inclination issue.
The noise in this galaxy was too high for the velocity dispersion to be 
measure, so we keep this fixed at $\sigma=10$\,km/s.

The results for the optically thin and the self-absorbing \hi models are 
shown in Figure \ref{fig:ESO138decomposition}.
The cross-correlation diagram of the optically thin model is shown in 
Figure \ref{fig:ESO138-crosscorr} (left), while the one for the self-absorption 
model is shown in Figure \ref{fig:ESO138-crosscorr} (right).
In both cases, the rotation curve and the vertical force gradient have 
been successfully fitted.
Only at the larger radii do observed and theoretical curves 
of the vertical force gradient start to deviate.

The optically thin model produces a halo core density 
$\rho_{0,\textrm{halo}}$ of $0.261^{+0.002}_{-0.003}$ M$_\odot$/pc$^3$ 
and a scale-length of $R_c =0.70^{+0.02}_{-0.01}$\,kpc.
The halo is distinctly prolate, with the optimal solution 
located at the boundary condition of $q =2.0^{+0.1}_{-0.2}$.
The stellar disc is very bright, with $M_*/L_R=0.50^{+0.02}_{-0.01}$.

Compared to this, the self-absorbing \hi model finds a halo 
with a higher core density of 
$\rho_{0,\textrm{halo}} = 0.620^{+0.009}_{-0.007}$\,M$_\odot$/pc$^3$ and 
a scale-length of $R_c  = 0.41_{-0.01}^{+0.02}$\,kpc.
Again, the optimal solution favors a prolate halo at $q =1.9^{+0.1}_{-0.3}$.
In this case the mass-to-light is $M_*/L_{R}= 0.69^{+0.04}_{-0.06}$.

\subsection{ESO\,274-G001}

Galaxy ESO\,274-G001 is one of the two galaxies from Paper III
for which we could accurately measure the velocity dispersion.
We model the galaxy in the standard way, setting the $M_*/L_{R}$ 
lower boundary at 0.5. 
Both the rotation curve and vertical force decomposition are shown 
in Figure \ref{fig:ESO274decomposition}, where we show the results 
for both the optically thin and self-absorbing \hi models.
Both models have reproduced  the rotation curve and the vertical 
force gradient reasonably well, although the self-absorbing model 
was more successful at the vertical force gradient.

We show the cross-correlation diagram for the optically thin \hi model in
Figure \ref{fig:ESO274-crosscorr} (left).
There is a clear correlation between the various parameters, which is mostly 
due to the uncertainty in $M_*/L_{R}= 0.96^{+0.46}_{-0.42}$.
The halo is oblate with 
$q =0.7^{+0.1}_{-0.1}$, $\rho_{0,\textrm{halo}} = 0.150^{+0.020}_{-0.022}$\,M$_\odot$/pc$^3$ 
and $R_c = 1.10^{+0.04}_{-0.03}$\,kpc.
The mass to light ratio $M_*/L_{R}$ was $0.96^{+0.46}_{-0.42}$.

The cross-correlation diagram of the \hi self-absorption model is shown in 
Figure \ref{fig:ESO274-crosscorr} (right).
There is again scatter in  $M_*/L_{R}$, although the value has dropped
 compared to the optically thin model. It is now at $0.76^{+0.57}_{-0.26}$.
The shape of the halo is identical to the optically thin model, with 
an oblate shape of $q=0.7^{+0.1}_{-0.1}$.
The other parameters are 
$\rho_{0,\textrm{halo}} = 0.094^{+0.009}_{-0.019}$\,M$_\odot$/pc$^3$ 
and $R_c = 1.42^{+0.09}_{-0.03}$\,kpc.
Compared to the optically thin model, the self-absorption \hi model produces a 
dark matter halo with longer scale length $R_c$ and lower central density 
$\rho_0$.

If we limit the analysis to $M_*/L_R<0.55$, we find that the halo becomes 
even more oblate, 
$q = 0.64_{-0.01}^{+0.01}$ for the optically thin and 
$q =0.67_{-0.01}^{+0.01}$ for the self-absorption model.
The central density of the halo also goes up to 
$\rho_0 = 0.171_{-0.001}^{+0.002}$\,M$_\odot$/pc$^3$ for the optically thin model, 
and 
$\rho_0 = 0.103_{-0.001}^{+0.001}$\,M$_\odot$/pc$^3$ for the self-absorbing \hi.
The scale lengths goes down to 
$h_0 = 1.07_{-0.01}^{+0.01}$\,kpc and 
$h_0 = 1.39_{-0.01}^{+0.01}$\,kpc respectively.
A lower mass in the stellar disc thus results in haloes that are more oblate, 
have higher central core densities and slightly shorter scale lengths.

\begin{figure*}
\centering
\includegraphics[width=0.40\textwidth]{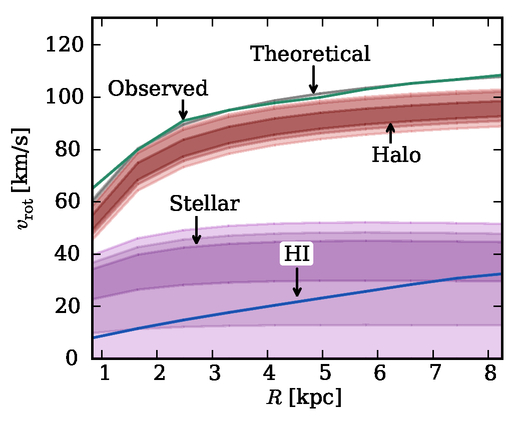}
\includegraphics[width=0.40\textwidth]{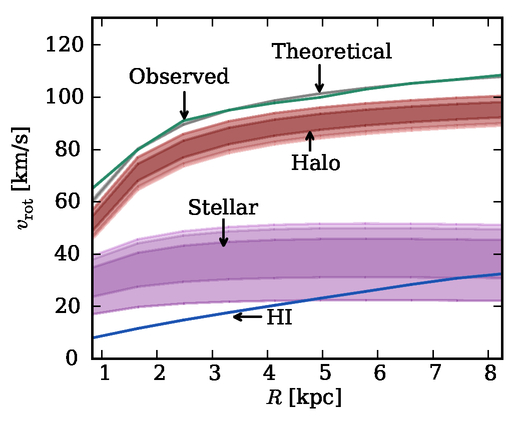}
\includegraphics[width=0.40\textwidth]{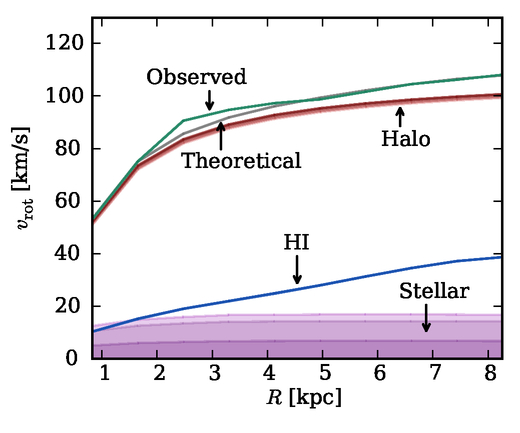}
\includegraphics[width=0.40\textwidth]{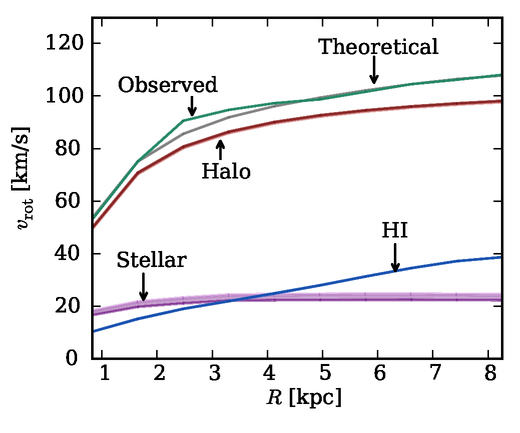}
\includegraphics[width=0.40\textwidth]{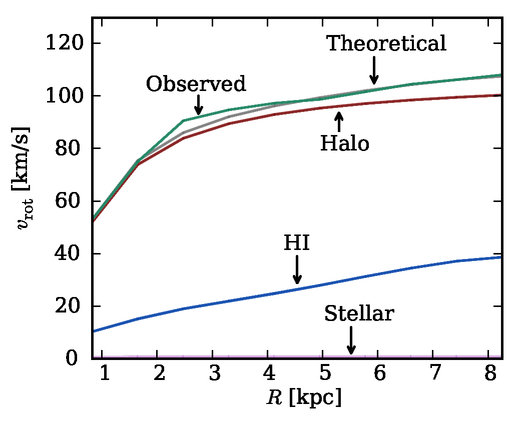}
\includegraphics[width=0.40\textwidth]{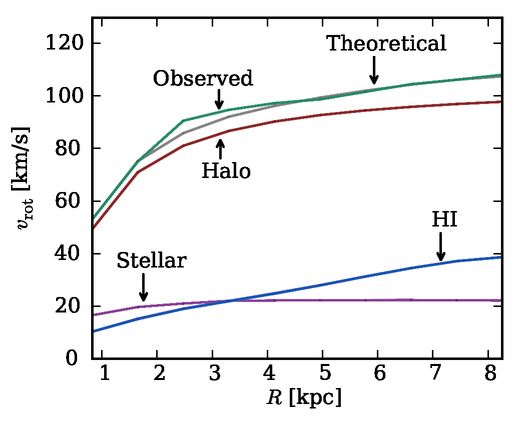}
\caption[Rotation curve decomposition for UGC\,7321]{Rotation curve decomposition of UGC\,7321 for various models. From top-left to bottom-right: optically thin \hi (M/L $\geq0$), optically thin \hi (M/L $\geq0.5$), self-absorption \hi (M/L $\geq0$),  self-absorption \hi (M/L $\geq0.5$), spherical halo with self-absorption \hi (M/L $\geq0$), spherical halo with self-absorption \hi (M/L $\geq0.2$)}\label{fig:UGC7321rotation}
\end{figure*}

\begin{figure*}
\centering
\includegraphics[width=0.40\textwidth]{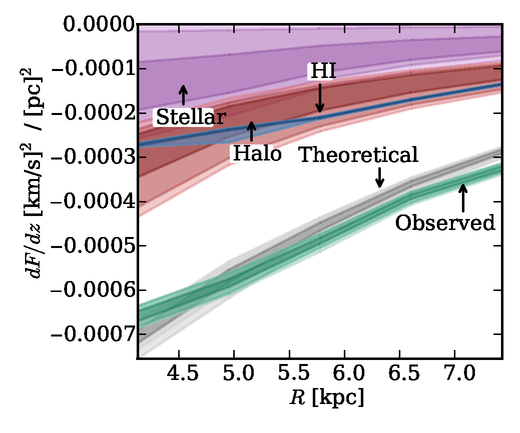}
\includegraphics[width=0.40\textwidth]{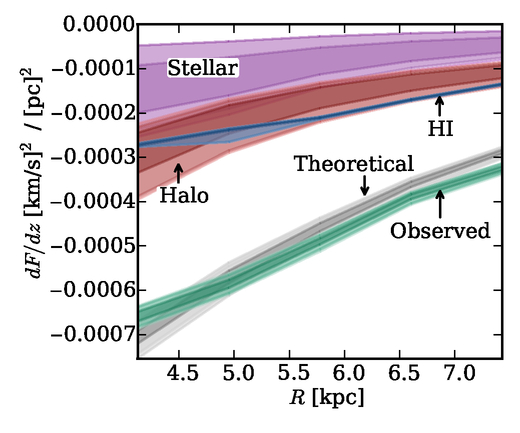}
\includegraphics[width=0.40\textwidth]{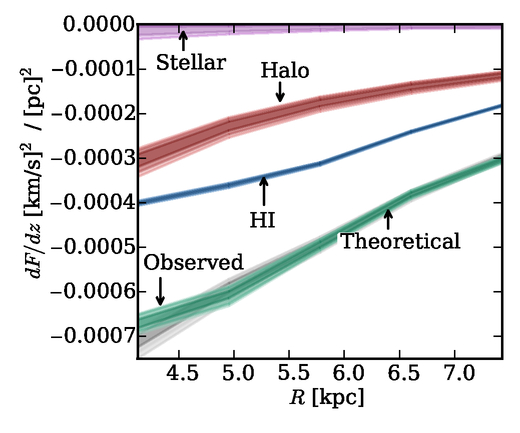}
\includegraphics[width=0.40\textwidth]{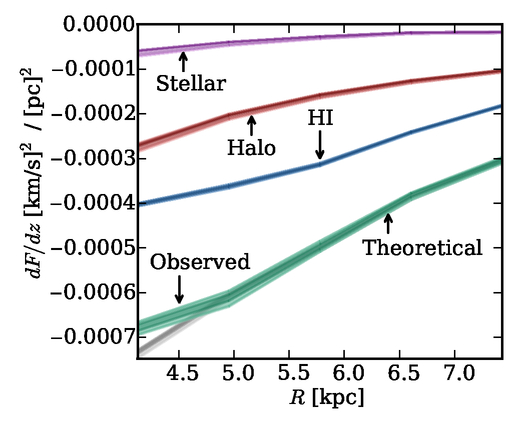}
\includegraphics[width=0.40\textwidth]{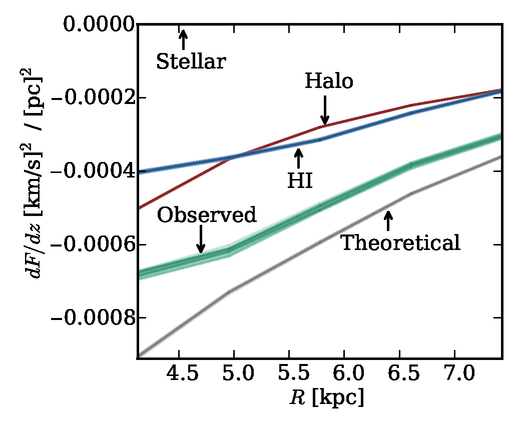}
\includegraphics[width=0.40\textwidth]{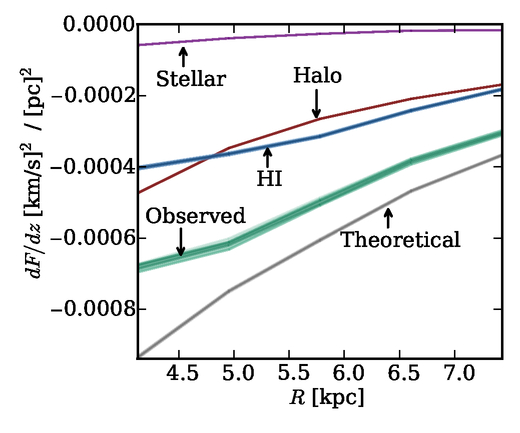}
\caption[Vertical force decomposition for UGC\,7321]{Vertical force decomposition of UGC\,7321 for various models. From top-left to bottom-right: optically thin \hi (M/L $\geq0$), optically thin \hi (M/L $\geq0.5$), self-absorption \hi (M/L $\geq0$),  self-absorption \hi (M/L $\geq0.5$), spherical halo with self-absorption \hi (M/L $\geq0$), spherical halo with self-absorption \hi (M/L $\geq0.5$)}\label{fig:UGC7321zforce}
\end{figure*}

\begin{figure*}
\centering
\includegraphics[width=0.49\textwidth]{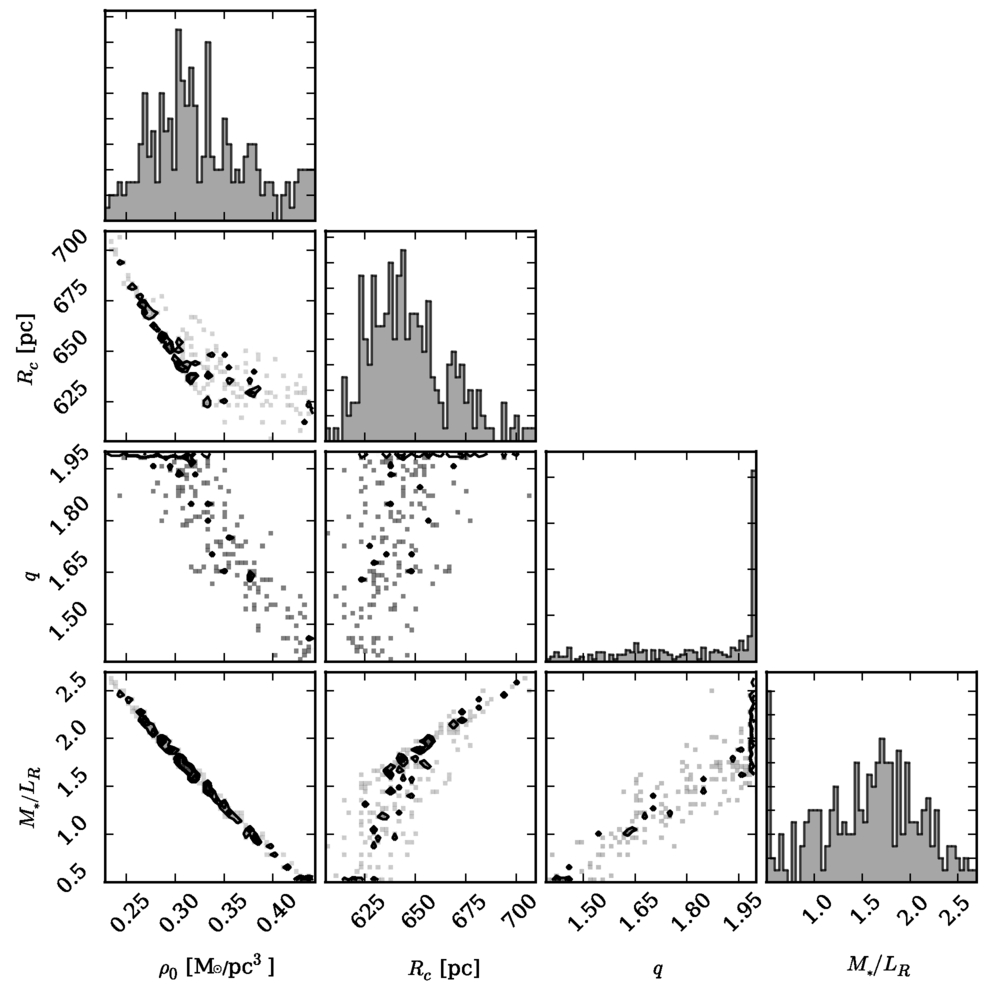}
\includegraphics[width=0.49\textwidth]{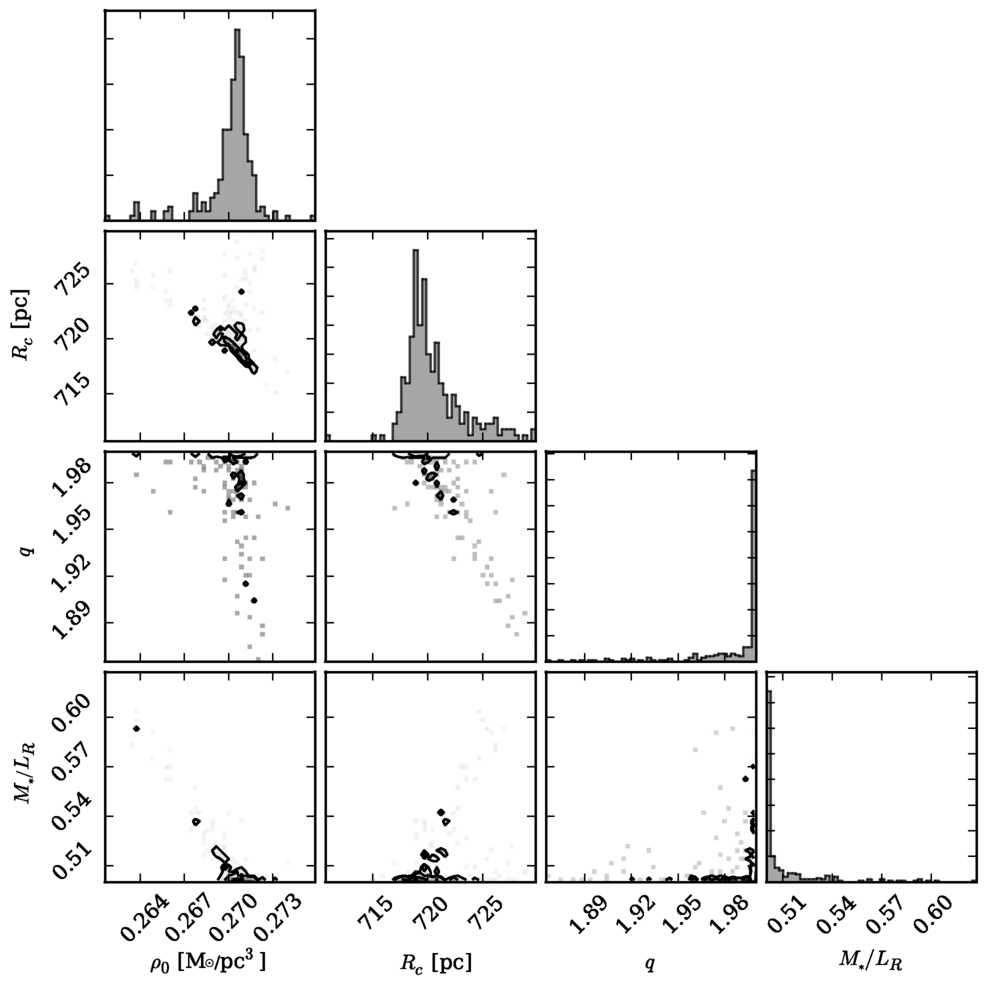}\caption[Halo parameters of UGC\,7321 (optically thin)]{Correlation diagram for the optically thin model of UGC\,7321 (left) and for the self-absorption model (right), both with a constraint of M/L$\geq0.5$.}\label{fig:UGC7321crosscorrelation}
\end{figure*}

\subsection{UGC\,7321}
Galaxy UGC\,7321 has the highest signal to noise ratio for the \hi 
data from our sample.
The galaxy was previously modelled by \citet{OBrien2010D}, who found 
that the halo flattening $q$ was round ($q=1.0\pm0.1$).
Their modelling strategy consisted of a two-pass scheme, in which 
they first performed a rotation curve decomposition, and only then 
performed a separate fit to the vertical force gradient.
This second fit however failed, and the authors had to drastically 
deviate from the results from the rotation curve decomposition, and 
use a very low mass stellar disc, in order to reproduce the observed vertical 
force gradient.
We have performed an inspection of the codes used by in the analysis of
\citet{OBrien2010D}. It appears there was a restriction which
allowed only models with $q\leq1$ and
 it would have been impossible for them to fit a prolate halo.

\citet{Banerjee2010A} also analysed {UGC\,7321} and found a spherical halo.
These authors assumed a constant velocity dispersion, or at most a 
decreasing gradient, in their work, and use a different potential than us.

In Figure \ref{fig:UGC7321rotation}, we demonstrate our own rotation 
curve decomposition of this galaxy, and in 
Figure \ref{fig:UGC7321zforce} we show the vertical force gradient 
decomposition.
Rather than present decompositions for only  the optically thin and 
self-absorbing \hi models that we measured, as for the previous 
galaxies, we show the results for six fits.

Since \citet{OBrien2010D} found a good fit at a negligible stellar mass, the 
first panel in both figures demonstrates a fit in which the $M_*/L_{R}$ can 
range between zero and three, for an optically thin \hi disc.
This fit should therefore be the closest to the results obtained by 
\citet{OBrien2010D}.
The 1000 samples produce a range of solutions. 
Both the rotation curve and the vertical force gradient are well reproduced.
The mass-to-light ratio has a median of $M_*/L_{R}= 1.58_{-0.68}^{+0.45}$, 
which is significantly higher than measured by \citet{OBrien2010D}. 
The halo is very prolate, $q =1.90_{-0.32}^{+0.10}$ and has a high central 
density of $\rho_{0,\textrm{halo}} = 0.324_{-0.043}^{+0.067}$\,M$_\odot$/pc$^3$ 
and short scale-length of $R_c = 0.64_{-0.02}^{+0.05}$\,kpc.
Previously, \citet{OBrien2010D} have reported $R_c = 0.52\pm0.02$\,kpc 
and $ \rho_{0,\textrm{halo}} =0.73\pm0.05$\,M$_\odot$/pc$^3$.

As we estimate a minimum of $M_*/L_{R}=0.5$, the second panel raises the 
boundary condition for the minimal stellar mass-to-light ratio to $0.5$.
The mass-to-light is found to be $M_*/L_{R}= 1.63_{-0.66}^{+0.48}$, still very 
similar to the previous model.
The observed rotation curve and vertical velocity gradients are  well 
reproduced, as shown in Figures \ref{fig:UGC7321rotation} and 
\ref{fig:UGC7321zforce}.
We find  $\rho_{0,\textrm{halo}} = 0.318_{-0.042}^{+0.064}$\,M$_\odot$/pc$^3$ 
and $R_c = 0.64_{-0.02}^{+0.03}$\,kpc, roughly similar values as the previous fit. 
The halo shape runs firmly towards $q=2$, which is also the boundary condition. 
We have tested the effect of lifting this boundary condition.
When we do this, the model tends to run towards even greater values of $q$.
However, since the current research question focuses primarily on prolate 
versus oblate, we have decided to stick to an upper boundary of $q=2$.
We present a cross correlation diagram of this fit in 
Figure \ref{fig:UGC7321crosscorrelation} (left).

Our next fit uses the self-absorption \hi model 
rather than the optically thin model.
We again let $M_*/L_{R}$ run from zero to three. 
As shown in Figures \ref{fig:UGC7321rotation} and 
\ref{fig:UGC7321zforce}, the stellar disc in this 
fit gets a negligible mass assigned ($M_*/L_{R}= 0.00_{-0.00}^{+0.04}$).
The other parameters are 
$\rho_{0,\textrm{halo}} =0.307_{-0.003}^{+0.002}$\,M$_\odot$/pc$^3$, $R_c = 0.71_{-0.01}^{+0.01}$\,kpc and $q=1.84_{-0.09}^{+0.12}$.
Compared to the optically thin model, the dark matter halo is again strongly 
prolate, but has a longer scale-length and lower central density.

Similar to the optically thin \hi case, we again increase the lower 
$M_*/L_{R}$ boundary to $0.5$.
The results are shown in Figures \ref{fig:UGC7321rotation} and 
\ref{fig:UGC7321zforce}.
The observed rotation curve has been modelled well, but the model fails to 
account for the vertical force gradient and produces too strong a vertical 
force gradient.
This directly illustrates why $M_*/L_{R}$ was zero in the previous fit, as 
this was the only way for the vertical force gradient to be fit.
The parameters found are 
$\rho_{0,\textrm{halo}} = 0.28_{-0.01}^{+0.01}$\,M$_\odot$/pc$^3$, $R_c =0.72_{-0.01}^{+0.01}$\,kpc, $q=2.00_{-0.03}^{+0.01}$ and $M_*/L_{R}=0.5$.
The cross-correlation diagram for this fit is shown in Figure 
\ref{fig:UGC7321crosscorrelation} (right).\label{sec:UGC7321-decomposition}

In their revious study, \citet{OBrien2010D} 
were unable to  fit the rotation curve and the vertical force gradient 
simultaneously. 
They successfully started with a  rotation curve decomposition, in which 
the stellar disc mass was a free parameter.
However, in order to subsequently perform their vertical force gradient 
decomposition, they were forced to drastically lower the stellar mass.
They eventually found a spherical halo, but only when they allowed very small
$M_*/L_{R}$ (the best fit was actually for $M_*/L_{R}=0$), smaller than we 
allowed here.

As two final tests, we ran fits to the optically thin and self-absorption 
\hi results, in which we constrained $q=1$.
The results are shown in Figures \ref{fig:UGC7321rotation} and 
\ref{fig:UGC7321zforce}.
The rotation curve decomposition does not depend strongly on $q$ 
\citep{OBrien2010D}. As such, it is again reproduced well.
Clearly, however, the vertical force gradient is fit poorly. 
A spherical halo simply does not work for this galaxy.

\begin{table*}
\begin{tabular}{lc|llll}
Name & \hi model&$\rho_0$ [M$_\odot$/pc$^3$] & $R_c$ [kpc] & $q$& $M_* / L_R$\\
\hline
IC\,5249&SA&$0.004^{+0.003}_{-0.001}$ & $9.79^{+0.21}_{-2.06}$ & $1.0^{+0.4}_{-0.3}$ &$2.98^{+0.02}_{-1.17}$\\
IC\,5249&OT&$0.007^{+0.005}_{-0.001}$ & $5.45^{+0.21}_{-0.52}$ & $1.5^{+0.5}_{-0.5}$ &$2.62^{+0.38}_{-1.19}$\\
ESO\,115-G021&SA&$0.015^{+0.001}_{-0.001}$ & $3.02^{+0.03}_{-0.08}$ & $0.5^{+0.1}_{-0.1}$ &$2.89^{+0.11}_{-0.44}$\\
ESO\,115-G021&OT&$0.022^{+0.004}_{-0.001}$ & $2.30^{+0.10}_{-0.15}$ & $0.7^{+0.1}_{-0.1}$ &$2.84^{+0.16}_{-1.15}$\\
ESO\,138-G014&SA&$0.620^{+0.009}_{-0.007}$ & $0.41^{+0.02}_{-0.01}$ & $1.9^{+0.1}_{-0.3}$ &$0.69^{+0.04}_{-0.06}$\\
ESO\,138-G014&OT&$0.261^{+0.002}_{-0.003}$ & $0.70^{+0.02}_{-0.01}$ & $2.0^{+0.1}_{-0.2}$ &$0.50^{+0.02}_{-0.01}$\\
ESO\,274-G001&SA&$0.094^{+0.009}_{-0.019}$ & $1.42^{+0.09}_{-0.03}$ & $0.7^{+0.1}_{-0.1}$ &$0.76^{+0.57}_{-0.26}$\\
ESO\,274-G001&OT&$0.150^{+0.020}_{-0.022}$ & $1.10^{+0.04}_{-0.03}$ & $0.7^{+0.1}_{-0.1}$ &$0.96^{+0.46}_{-0.42}$\\
UGC\,7321&SA&$0.270^{+0.001}_{-0.001}$ & $0.72^{+0.01}_{-0.01}$ & $2.0^{+0.1}_{-0.1}$ &$0.50^{+0.02}_{-0.01}$\\
UGC\,7321&OT&$0.318^{+0.064}_{-0.042}$ & $0.64^{+0.03}_{-0.02}$ & $1.9^{+0.1}_{-0.3}$ &$1.63^{+0.48}_{-0.66}$\\
\hline
\end{tabular}
\caption[Measured parameters for the various haloes]{Measured parameters for the various haloes. OT denotes the optically thin \hi models, while SA denotes the self-absorbing \hi models.}
\label{tbl:DMhaloes}
\end{table*}

In the previous section, we have presented the results for the individual 
galaxies. 
So how do the results compare to each other? 
In Table \ref{tbl:DMhaloes}, we present an overview of all the derived 
parameters. 
For ESO\,138-G014, we only present the results where the thickness of 
the \hi layer has been reduced by 30\%.
For galaxy UGC\,7321, we present the results for the default model, in 
which the mass-to-light ratio $M_*/L_R$ was allowed to vary between 0.5 
and 3.0, and the halo shape $q$ between 0.1 and 2.0.

We present an overview of the average of these parameters in Table 
\ref{tbl:globalhalo}. 
There is an interesting difference between optically thin and 
self-absorbing \hi models. 
Overall, we see that the halo of an optically thin \hi model 
has a core density that is overestimated by $150\%$.
The scale length of the dark matter halo is $28\%$ longer in 
the self-absorption model, compared to the optically thin model. 
In addition, where the optically thin models have a median shape 
that is prolate, the median shape is spherical in self-absorption \hi models.
The mass-to-light ratio of the stellar disc drops by more than 
half when self-absorbing \hi is accounted for. 

All of our five discs are sub-maximal.
This was already demonstrated by \citet{OBrien2010D} and 
\citet{Banerjee2010A} for UGC\,7321, who reported a stellar 
disc M/L$_R$ at a maximum of 2.5, although their final 
decomposition found a maximum of M/L$_R$ = 0.2. 
The result is consistent with the work by \citet{Bershady2011A}, 
who argued that all galaxies are sub-maximal based on an analysis 
of the central vertical velocity dispersion of the discs stars and 
the maximum rotation of 30 face-on galaxies.
Similar conclusions have previously been reached by \citet{bot97} 
and \citet{Kregel2005}.
\citet{Martinsson2013A} confirmed these results after performing 
dynamically determined rotation curve mass decompositions for all 
these 30 galaxies.

\begin{table*}
\centering
	\begin{tabular}{c|llll|l}
	Name&$\rho_0$ [M$_\odot$/pc$^3$] & $R_c$ [kpc] & $q$& $M_* / L_R$ & Citation\\
	\hline\hline
	UGC\,7321 & $0.73\pm0.05$& $0.52\pm0.02$&$1.0\pm0.1$ & $0.2$ &\citet{OBrien2010D}\\
	UGC\,7321 & $0.048\pm0.009$ & $2.7\pm0.2$& 1&  &\citet{Banerjee2010A}\\
	NGC\,4244 & $0.6595\pm0.6495$ & $6.15\pm5.85$ & 0.2 & & \citet{Olling1996A}\\
	M\,31  &  0.011 & 21 & 0.4 & & \citet{Banerjee2008A}\\
	Galaxy (NFW) & $0.0413_{-0.016}^{+0.013}$ & $9.26_{-4.2}^{+5.6}$ & && \citet{Nesti2013}\\
	Galaxy & $0.0147_{-0.0093}^{+0.029}$ & $16.1_{-7.8}^{+17}$ && & \citet{Nesti2013}\\
	Galaxy & & 12 & 5/3 & & \citet{Helmi2004A}\\
	Galaxy & & 7.1 & 0.7 & & \citet{Olling2000}\\
	Lensing & & $23.9_{-0.5}^{+0.2}$ & $0.62_{-0.25}^{+0.25}$& &\citet{vanUitert2012}\\
	
	ESO\,138-G014 (NFW)&$0.013\pm0.002$&$7.5\pm0.5$& & & \citet{Hashim2014}\\
	ESO\,138-G014 (Burkert)&$0.077\pm0.042$&$10.7\pm2.7$& & & \citet{Hashim2014}\\
	\hline\hline
	\end{tabular}
\caption{Overview of haloes measured in other papers that were used in Figure \ref{fig:DMdiscussion} as data points.}\label{tbl:otherDM}
\end{table*}

\begin{figure}
\centering
\includegraphics[width=0.23\textwidth]{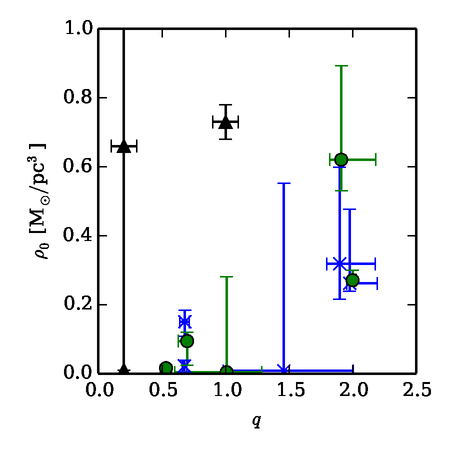}
\includegraphics[width=0.23\textwidth]{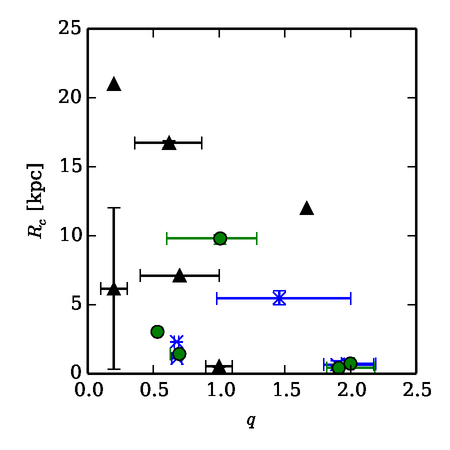}
\includegraphics[width=0.23\textwidth]{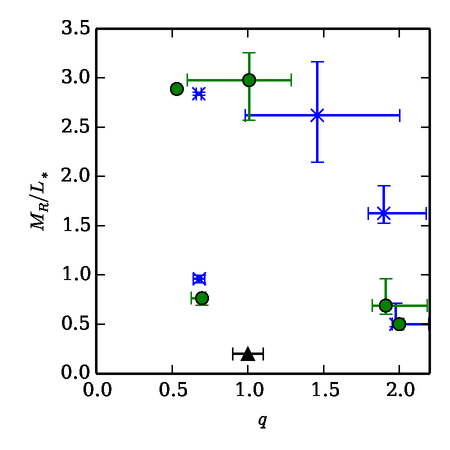}
\includegraphics[width=0.23\textwidth]{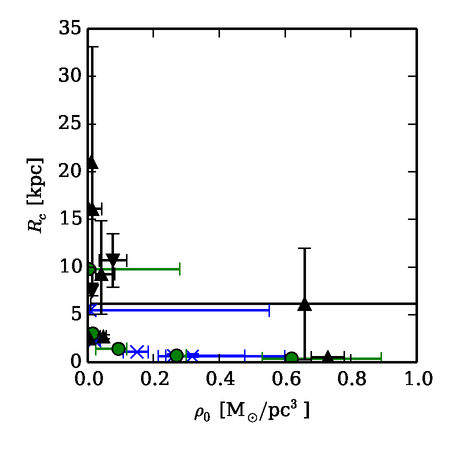}
\includegraphics[width=0.23\textwidth]{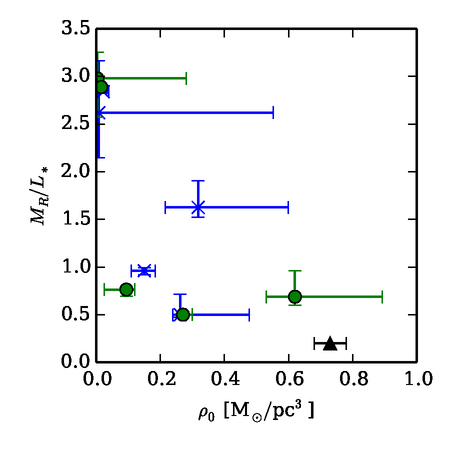}
\includegraphics[width=0.23\textwidth]{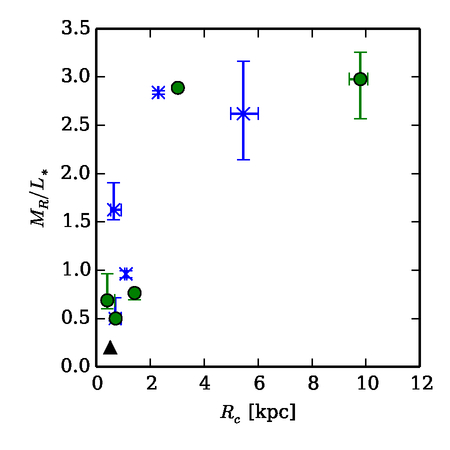}
\caption[Overview of parameters]{Overview of the correlations between the various halo parameters. Blue crosses represent the halo parameters for the self-absorbing \hi, while green circles represent the optically thin models. Black markers are the data points from Table \ref{tbl:otherDM}}\label{fig:DMdiscussion}
\end{figure}

In Figure \ref{fig:DMdiscussion}, we demonstrate the correlation between the 
four free parameters from our fit. 
We have also included a range of points from other authors in this view 
(see Table \ref{tbl:otherDM} for an overview, note that multiple halo 
models are used an as such the core radius can be expected to vary).
Inspecting the figure, the most notable correlations is the one between 
core radius $R_c$ and halo core density $\rho_0$. 
With the exception of one point, ESO\,138-G14 by \citet{Hashim2014} using 
a NFW halo, all of the points seem to follow a relation of 
$R_c\approx1/\rho_0$. Our result for ESO\,138-G14 are uncertain due to
the possiblity of a r4esisual inclination compared to edge-on and any results
on this galaxy, including ours at $q\sim 2$ should be treated with caution. 
This relation is similar to the degeneracy between the two parameters in an 
individual galaxy, as for example in Figure 11, and 
it interesting to observe a similar trend visible across multiple galaxies 
and halo models. 
If this is a true relation, then it implies that there are two families of 
haloes: one compact halo family with high core density $\rho_0$ and scale 
length $R_c$, and a second non-compact halo family with low core density 
$\rho_0$ and scale length $R_c$.

\begin{table}
\centering
\begin{tabular}{c|cc}
name & optically thin & self absorbing\\
\hline\hline
$\rho_0$ &$0.150\pm 0.124$& $0.094 \pm0.230$\\
$R_c$ & $1.10\pm 1.81$ &$1.42\pm 3.48$\\
$q$  & $1.5\pm 0.6$ &$1.0\pm 0.6$\\
$M_*/L_R$  &$1.63\pm 0.91$ &$0.76\pm 1.12$\\
\hline\hline
\end{tabular}
\caption[Global halo parameters]{Overview of the global parameters of our halo sample. Shown are the median and standard deviations of the parameters. The units of $\rho_0$ are in M$_\odot$/pc$^3$ and those of the radius $R_c$ in kpc.}
\label{tbl:globalhalo}
\end{table}

Our best results are for ESO\,274-G001 and UGC\,7321, where we have been able 
to include a measurement of the velocity dispersion of the \hi as function 
radius in the analysis. 
Focusing on these two, we find two very different haloes. 
In ESO\,274-G001, the halo is oblate with a shape $q=0.7\pm0.1$ (regardless 
of the \hi model); while in UGC\,7321, the halo is distinctly prolate with 
a shape of $q=1.9^{+0.1}_{-0.3}$ (optically thin \hi) and $q=2.0\pm0.1$ 
(self-absorbing \hi).

So how do these shapes compare to other galaxies?
Looking at our own Galaxy, \citet{Law2010A} propose a triaxial dark matter 
halo for the Milky Way, in which $(c/a)_\phi=0.72$ and $(b/a)_\phi=0.99$. 
\citet{Banerjee2011A} proposed a dark matter halo shape for the Galaxy  
which becomes progressively more prolate with radius. 
 \citet{VeraCiro2013} report an oblate halo with $q=0.9$ for the inner 
10\,kpc, based on stellar streams.
Using lensing, \citet{Barnabe2012A} also found a slightly oblate halo at 
$q=0.91^{+0.15}_{-0.13}$ for galaxy SDSS\,J2141.
The large weak lensing galaxies sample of \citet{vanUitert2012}, in which 
$2.2\times10^7$ galaxies were studied, produced a halo ellipticity 
distribution that also favors oblate haloes. 
The distributions of the halo shape $q$ was $0.62^{+0.25}_{-0.25}$. 
The three polar ring galaxies studied by \citet{Whitmore1987A}, A0136-0801, 
NGC\,4650A and ESO\,415-G026, had slightly oblate to spherical halo shapes 
$q$: $0.98\pm0.20$, $0.86\pm0.21$ and $1.05\pm0.17$. 
From this selected sample of papers, it becomes apparent that our result for 
ESO\,274-G001, with $q=0.7\pm0.1$, is consistent with other papers. 

While ESO\,274-G001 clearly matches up with measurements in other galaxies, 
the other case of our two best fits, UGC\,7321, is a more `problematic' one 
(Figures 13, 14, 15 and 16). 
With a halo shape of  $q=1.9_{-0.3}^{+0.1}$ for the self-absorption model, the 
dark matter halo shape is very strongly prolate.
As we commented before in Section \ref{sec:UGC7321-decomposition}, our upper 
boundary condition for the halo shape is $q\leq2$. If we had removed this 
boundary, some of the fits returned results as high as $q\sim5000$, which 
are clearly not physical.
The galaxy has been previously analysed by \citet{Banerjee2010A}, whom 
successfully modelled the dark matter halo shape for a spherical halo.
\citet{OBrien2010D} had problems fitting the dark matter halo shape. 
They had to lower their initially measured asymptotic halo rotation (see 
Equation \ref{eqn:dmhalovelocity}) in order to get a successful fit to their 
data at $q=1.0\pm0.1$, although they were limited to $q\leq1$ in their analysis.
Had their boundary condition been higher, it would have been likely that they 
too would have found higher values for $q$.

\subsection{Concerns regarding reliability and degeneracy}
Given that our two best galaxies produce such drastically different results, 
how reliable is our methodology? 
To answer this question, let us recap the underlying assumptions from this 
paper and Paper IV.
\subsubsection{Concerning the neutral hydrogen}

We start  with the neutral hydrogen.
In Paper I, we argued that the \hi in edge-on galaxies 
could suffer from significant self-absorption. 
To model the \hi more accurately, we developed a new tool that allowed 
the neutral hydrogen in galaxies to  be fit automatically, while 
incorporating a treatment for the self-absorption of the gas.
Indeed, we saw in Section 7 of Paper II that the 
visible mass of a galaxy drops as one rotates it from face-on to 
fully edge-on. 
In Paper II, we developed a method 
to model the \hi content of a galaxy that was edge-on. 
In Section 3 of Paper III, we tested this method on 
a series of simulated galaxies, showing that we could reproduce the 
input parameters reasonably well using our method. 
We also demonstrated that assuming an optically thin \hi disc, which 
in reality was self-absorbing, could lead to a wrong measure of face-on 
surface density, thickness of the \hi layer and the velocity dispersion.
We have continued the use of the optically thin \hi results into this 
paper to demonstrate how the dark matter halo measurement is affected by this.
As discussed in the previous section, the results are  drastic.
How accurate are our results now?

One of the key assumptions made in Paper II 
was the effective spin temperature of the neutral hydrogen of $100$\,K. 
While this was has proven a very successful value on which to base our 
results, it is an assumption based purely on what seemed to work best.
In reality the neutral hydrogen most likely consists of multiple phases, such 
as the cold neutral medium (CNM) and warm neutral medium (WNM). 
The effective spin temperature is a result of the mix of the phases of the 
CNM, which has a median spin temperature of 80\,K and the WNM, with 
temperatures between 6000 and 10000\,K.
In Section 4 of Paper II, we demonstrated how the interplay 
of \hi gas phases could lead to an effective median spin temperature.
So what would be the consequence of a wrong estimate of the spin temperature? 
Suppose that the spin temperature would have been $T_\textrm{spin}=90$\,K 
rather than 100\,K. 
In that case the face-on surface density of the neutral hydrogen will 
be higher, which in this paper would lead to higher theoretical 
rotation from the gas components in the galaxy and stronger vertical 
force gradients then are currently found.
Simultaneously, the thickness of the disc would be smaller, and thus 
the total observed vertical force gradient would be larger (Equation 
\ref{eqn:Ftotal}).
Although hard to estimate the exact effect, the phases of the \hi all 
have different distributions, together producing the observed thickness 
of the disc \citep{Lockman1991}. 
The effective spin temperature could thus be height dependent as well.

Another assumption is the uniform density of the \hi as a function of radius.
In reality of course, galaxies have spiral arms, supernova, shocks, 
gravitational collapse, and other features, all of which create a 
drastically non-uniform \hi disc. 
The question thus remains how strongly the parameters are affected by this. 
\citet{Kamphuis2013A} made a valiant attempt to model the density waves 
in galaxies NGC\,5023 and UGC\,2082, demonstrating that these could be 
detected in edge-on galaxies\footnote{These authors did not 
model the \hi as self-absorbing, which most likely is hampering their results.}.
Indeed, as we discussed in Section 5 of Paper I, the position-velocity 
(XV-)diagrams are not symmetric on both side of the galaxies.
This problem would most strongly affect the velocity dispersion, which is 
dependent upon small-scale features. 
In most cases, this leads to an overestimation of the \hi velocity dispersion, 
as the fitting algorithm tries to `smooth over' the small-scale fluctuations.
The result would be an overestimation of the observed 
vertical force gradient through Equation \ref{eqn:Ftotal}. 
This could be a likely reason why UGC\,7321 has such a distinctive halo shape.

We have also assumed that the \hi has an isotropic velocity dispersion 
tensor, i.e has the same value in the $R$, $z$ and $\theta$ directions 
(see Section 2.2 of Paper II).
While there currently is no observational proof that this is an invalid 
assumption, it remains untested.
If the velocity dispersion tensor were in reality anisotropic, it would imply 
that our observed vertical force gradient is wrong (Equation \ref{eqn:Ftotal}). 
Simultaneously, it would affect the amount of observed \hi in the 
self-absorption mode, as gas with low velocity dispersion would give
a alrger effect than gas with high velocity dispersion 
(Equation 19 of Paper II).
It would affect the rotation curve measurements to a minor degree.

We also have assumed that the velocity dispersion is isothermal 
in $z$, i.e. does not vary with height. 
In Section 4.7 of Paper III, we have attempted to 
measure this in ESO\,274-G001. 
We found a very small increase of 1\,km/s in the slice above the central 
290\,pc of the disc. 
If this were confirmed in other galaxies, it would mean that Equation 
\ref{assumption:isothermal} is false, and thus our observed total force 
gradient would have been wrong.
Previously, \citet{Lockman1991} has attempted to model the vertical 
structure of the Galaxy using multiple \hi phases, which had different 
scale heights and velocity dispersions. 
The phases with the highest scale heights also have the highest velocity 
dispersions.
It is quite possible that the effective spin temperature of the different
phases would increase for those with larger scale heights.
Assuming that our currently observed velocity dispersion is due to the 
combination of high and low $z$ gas,  the total vertical force gradient 
than we are currently reporting would be weaker in the mid-plane, and 
stronger at high values of $z$.

The thickness of the \hi disc has been assumed to follow a Gaussian form 
(see Equation 15 of Paper II) for mathematical convenience in 
Equation \ref{eqn:Ftotal}.
Other possible candidates for the \hi model would have been the sech 
and sech$^2$ functions.
\citet{Olling1995A} previously discussed the various types of discs and 
concluded that the changes due to this would be minor.
A sech function has more extended wings and  steeper inner slopes, 
compared the Gaussian function.
If our galaxies have high-projected latitude gas, such as due to warps 
or \hi haloes, then it is possible that a fit with a Gaussian function 
would find the FWHM of the disc to be unrealistically large.
A fit with a sech or even sech$^2$ function could then be a better 
approximation to the shape of the \hi disc. 

Another assumption in our model is the perfect edge-on nature of the \hi disc. 
In Section 3.5 of Paper III, we tested how our \hi fitting strategy 
worked on a galaxy at $i=88.8^\circ$.
We found that the parameters were well recovered. 
However, suppose that some galaies are even further from edge-on than that, as 
formally indicated by our stellar decompositions (Table 4 of Paper IV).
In that case, the thickness of the \hi is probably overestimated and the 
circular rotation underestimated. 
This has a profound effect on the rotation curve decomposition, which 
would require a larger rotational contribution due to the dark matter.
In a similar vein, the total observed vertical force gradient would be 
underestimated, and would thus require a more massive dark matter halo in 
the decomposition. 
On the other hand, the density of the \hi would be lower at a height of 
100\,pc, such that in Equation \ref{eqn:poissonz} the vertical force 
gradient due to the \hi would be lower.

\subsubsection{Concerning the stellar disc}
In Paper IV, we set out to use the \textsc{FitSKIRT} 
tool to model the stellar disc of our sample of galaxies.
We modelled the galaxies using a stellar disc and a bulge component.
While the  results were acceptable in most cases, these galaxies were 
selected to be relatively bulge-less (Section 3 of Paper I). 
Because of this, the fitting routine was found to `misuse' the bulge 
component as a tool to better model the stellar disc.
Thus, the bulges in most of the galaxies serve more like an extension 
of the stellar disc, rather than like a separate central component (see 
Table 4 of Paper IV for the parameters).
In some cases, the amount of light from the bulge component is similar 
to that of the stellar disc itself (Table 3 of Paper IV). 
Due to the massive amount of processing power required to perform the 
stellar decompositions, we have been unable to test how well the fitting
of just an exponential disc to the data would have worked.
Most likely, the results are similar to the combined parameters adopted here.

The stellar models in Table 4 of Paper IV were often found to 
deviate from complete edge-on. 
In the absence of dust lanes, which could prove an independent check, it 
remains unclear how accurate this result is.
If the galaxies were in reality more edge-on than measured, the stellar 
discs would have shorter scale heights. 
In Equation \ref{eqn:poissonz}, this would imply a lower stellar density 
at a height of $z=100$\,pc, and thus would require a stronger dark matter 
halo vertical force gradient.

\subsubsection{Concerning the cross-correlation between parameters}
An advantage of our MCMC method is that we have not one, but a whole 
range of parameter sets for each galaxy. 
This allows us to explore the interplay between the cross-correlations, 
such as demonstrate Figure 12 for the 
self-absorption model of ESO\,274-G001.
As is clear from this figure, a whole range of solutions can be valid. 
For example, one parameter set returned a core density of
 0.1\,M$_\odot$/pc$^3$, with scale length of 1.4\,kpc, halo 
shape $q$ of $0.67$ and $M_*/L_R$ of 0.4. 
A different but equal parameter set returned a core density of 
$0.06$,M$_\odot$/pc$^3$, with scale length of nearly 1.6\,kpc, shape 
$q=0.84$ and $M_*/L_R$ of 1.7.
These results are drastically different, yet both are accepted parameter sets. 

The largest source of uncertainty is the stellar disc $M_*/L_R$. 
It was beyond the scope of this project to measure this parameter 
in each of our galaxies, which is why we have adopted it as a free parameter.
A different solution would have been to adopt the maximum value of 
$M_*/L_R$ permitted in the rotation curve decomposition, the so-called 
maximum disc approach (e.g. \citet{Carignan1985,vanAlbada1985A}). 
However, these galaxies were selected to be dark matter dominated at all 
radii and as such this approach would have been invalid 
(Section 3 of Paper I).
In addition, the applicability of the maximum disc criteria has 
already been questioned  by \citet{Kregel2005} and \citet{Martinsson2013}, 
whom both report sub-maximal stellar discs.
Although beyond the scope of this project, the best approach would be to 
perform a full stellar population synthesis analysis of each galaxy. For 
examples see \citet{Bruzual2003} and \citet{Maraston2005}.
By measuring M$_\odot$/pc$^3$ rather than using it as a free parameter, the 
solution space becomes far less degenerate and the parameters can thus be 
fixed far more accurately.

Another cause of concern is the boundary conditions imposed upon our data. 
We have done our best to impose realistic boundary conditions.
For the dark matter halo shape, we adopted $0.1<q<2.0$, as we believe that 
even more oblate or prolate haloes would be unrealistic.
In Section \ref{sec:MLdiscussion} we adopted $0.5<M_*/L_R<3.0$ as a likely 
boundary, based on the 
stellar population models by \citet{Worthey1994A} and \citet{Bertelli1994A}.
As can clearly be seen from the various cross-correlation diagrams, the 
models often still run into the boundary conditions.
While it is possible to raise or remove the boundary conditions, we do not 
believe that this would lead to realistic results and we have therefore 
refrained from doing so.

\subsubsection{Halo model}
In this work, we have adopted the dark matter halo model by 
\citet{Sackett1994A}.
With this model, we can create  flattened, axi-symmetric, 
pseudo-isothermal haloes. 
In this model, the density is stratified in concentric ellipsoids.
We have chosen this model to be able to compare directly 
to \citet{OBrien2010D}, who also use this model.

There are many other halo model. 
\citet{Carignan1985, Carignan1988} used isothermal, rather 
than pseudo-isothermal haloes to model their galaxies.
\citet{Kormendy2004IAU} compared the merits between isothermal 
and pseudo-isothermal haloes. 
As is shown in that paper 
\citep[and reproduced in][]{OBrien2010D}, the rotation curve 
of an isothermal halo initially rises above the asymptotic 
velocity $v_H$, before dropping towards it again.
In contrast, the pseudo-isothermal rotation curve approaches 
the asymptotic velocity gradually from lower values.
This behavior would affect the results for the rotation 
curve decomposition.

There are more models, such as the NFW and Burkert halo model, 
each of which has some mathematical or theoretical advantage 
\citep{Burkert1995,Navarro1996}.
Even more exotic models exist in which the dark matter halo shape 
can vary with radius \citep{VeraCiro2013}. 
This can for example lead to haloes that get progressively more prolate 
with radius \citep{Banerjee2011A}.
While all these haloes are very interesting, we believe that the quality 
of the data, as discussed in this section, does not warrant such a 
detailed exploration of the properties of the various halo models. 

An altogether different solution would have been the use of MOND, which
 would have removed the need for a dark matter halo altogether 
\citep{Milgrom1983A}.
We find that in many of our vertical force decompositions, a 
slight increase in the mass of the \hi and stellar disc would 
be sufficient to account for the total observed vertical force gradient. 
As we argued before, additional mass in both the stellar and 
\hi discs is allowed for by the data. 
While it is beyond the scope of this project to test MOND on our data, it is 
an interesting avenue for further research.

\section{Conclusions}\label{sec:hydrostaticssummary}
We have attempted to measure the shape of the dark matter halo in five 
galaxies, using a simultaneous decomposition of the
rotation curve and of the vertical force gradient at the mid-plane.
For a dark matter halo model, we have adopted the \citet{Sackett1994A} 
dark matter halo.
Both optically thin and self-absorbing \hi models were used. 
We find that this leads to drastically different results.
As we have argued in Papers I, II and III, the \hi self-absorption models are 
the more accurate representation of galaxies.
Using \hi self-absorption, we found that a typical dark matter halo has 
a less dense core ($0.094\pm0.230$\,M$_\odot$/pc$^3$)\footnote{Central 
value is the median, error is the standard deviation.} compared 
to an optically thin \hi model ($0.150\pm0.124$\,M$_\odot$/pc$^3$).
The \hi self-absorption dark matter halo had a longer scale length $R_c$ of 
$1.42\pm 3.48$\,kpc, versus $1.10\pm 1.81$\,kpc for the optically thin \hi 
model.
The median halo shape was spherical, at $q=1.0\pm0.6$ (self-absorbing), while 
it was prolate at $q=1.5\pm0.6$ for the optically thin.

Our best results were obtained for ESO\,274-G001 and UGC\,7321, for which we 
were able to measure the velocity dispersion in Paper III. 
These two galaxies have drastically different halo shapes.
ESO\,274-G001 was found to be oblate at $q=0.7\pm0.1$ (both 
\hi models), while UGC\,7321 returns a distinctly prolate halo 
at $q=1.9_{-0.3}^{+0.1}$ (optically thin) and $q=2.0\pm0.1$ (self-absorbing). 
The halo of ESO\,274-G001 iss similar to those found in other studies, but 
UGC\,7321 is more problematic. 
In UGC\,7321, the most likely cause of concern is the presence of spiral 
arms and an \hi halo.

With these drastically different results, we concluded that the question 
whether haloes are oblate or prolate is not settled. 
The results for both of our best galaxies appear to be fine. A  larger 
set of galaxies needs to be analysed, before it can become clear if one 
of these galaxies is an outlier, or if prolate and oblate haloes are 
equally likely in nature.

We extensively discussed the various assumptions and sources of 
uncertainty in our models, of which there are many.
While we have done our best to treat for these assumptions, for 
example using MCMC fits to the \hi cube, we found that fitting the 
hydrostatics of the dark matter halo using the vertical force gradient 
near the mid-plane of the galaxy will always be tricky.

\section*{Acknowledgments}
SPCP is grateful to the Space Telescope Science Institute, Baltimore, USA, the 
Research School for Astronomy and Astrophysics, Australian National University, 
Canberra, Australia, and the Instituto de Astrofisica de Canarias, La Laguna, 
Tenerife, Spain, for hospitality and support during  short and extended
working visits in the course of his PhD thesis research. He thanks
Roelof de Jong and Ron Allen for help and support during an earlier 
period as visiting student at Johns Hopkins University and 
the Physics and Astronomy Department, Krieger School of Arts and Sciences 
for this appointment.

PCK thanks the directors of these same institutions and his local hosts
Ron Allen, Ken Freeman and Johan Knapen for hospitality and support
during many work visits over the years, of which most were 
directly or indirectly related to the research presented in this series op 
papers.

Work visits by SPCP and PCK have been supported by an annual grant 
from the Faculty of Mathematics and Natural Sciences of 
the University of Groningen to PCK accompanying of his distinguished Jacobus 
C. Kapteyn professorhip and by the Leids Kerkhoven-Bosscha Fonds. PCK's work
visits were also supported by an annual grant from the Area  of Exact 
Sciences of the Netherlands Organisation for Scientific Research (NWO) in 
compensation for his membership of its Board.


\bibliography{refsV}
\bibliographystyle{mn2e}


\bsp

\label{lastpage}

\end{document}